\documentclass[%
reprint, 
superscriptaddress, 
amsmath,amssymb,mathrsfs,amsfonts
aps,
]{revtex4-2}
\usepackage{amsmath,amssymb,amsfonts}
\usepackage{mathbbol}
\usepackage{float}
\usepackage{subfig}
\usepackage{physics}
\usepackage{graphicx}
\usepackage{dcolumn}
\usepackage[normalem]{ulem}
\usepackage[
hidelinks,
letterpaper,
pagebackref=false,
bookmarksopen,
bookmarksnumbered,
]{hyperref}
\usepackage{booktabs}
\usepackage{mathrsfs}

\newcommand{\twonorm}[2]{
	\norm{#1}_{\text{TS, }#2}
}
\newcommand{\margin}{\mathcal{M}}
\newcommand{\opnorm}[1]{\norm{#1}_{\text{op}}}

\newcommand{\Prod}{\mathrm{Prod}}
\usepackage{cleveref}
\usepackage{soul}
\usepackage[authormarkuptext=name,commentmarkup=uwave, highlightmarkup=uuline]{changes}
\usepackage[justification=raggedright, singlelinecheck=false]{caption}
\definechangesauthor[name={E},color=blue]{Eli}
\definechangesauthor[name={A},color=red]{Avshalom}
\definechangesauthor[name={R},color=olive]{Ron}

\newcommand{\addedsharon}[2][]{%
	\ifx&#1&%
	\added[id=Avshalom]{#2}%
	\else
	\added[id=Avshalom, comment=#1]{#2}%
	\fi
}
\newcommand{\highlightsharon}[2][]{%
	\ifx&#1&%
	\highlight[id=Avshalom]{#2}%
	\else
	\highlight[id=Avshalom, comment=#1]{#2}%
	\fi
}
\newcommand{\deletedsharon}[2][]{%
	\ifx&#1&%
	\deleted[id=Avshalom]{#2}%
	\else
	\deleted[id=Avshalom, comment=#2]{#1}%
	\fi
}
\newcommand{\replacedsharon}[3][]{%
	\ifx&#1&%
	\replaced[id=Avshalom]{#3}{#2}%
	\else
	\replaced[id=Avshalom, comment=#1]{#3}{#2}%
	\fi
}

\newcommand{\addedron}[2][]{%
	\ifx&#1&%
	\added[id=Ron]{#2}%
	\else
	\added[id=Ron, comment=#1]{#2}%
	\fi
}
\newcommand{\highlightron}[2][]{%
	\ifx&#1&%
	\highlight[id=Ron]{#2}%
	\else
	\highlight[id=Ron, comment=#1]{#2}%
	\fi
}
\newcommand{\deletedron}[2][]{%
	\ifx&#1&%
	\deleted[id=Ron]{#2}%
	\else
	\deleted[id=Ron, comment=#2]{#1}%
	\fi
}
\newcommand{\replacedron}[3][]{%
	\ifx&#1&%
	\replaced[id=Ron]{#3}{#2}%
	\else
	\replaced[id=Ron, comment=#1]{#3}{#2}%
	\fi
}

\newcommand{\addedeli}[2][]{%
	\ifx&#1&%
	\added[id=Eli]{#2}%
	\else
	\added[id=Eli, comment=#1]{#2}%
	\fi
}
\newcommand{\highlighteli}[2][]{%
	\ifx&#1&%
	\highlight[id=Eli]{#2}%
	\else
	\highlight[id=Eli, comment=#1]{#2}%
	\fi
}
\newcommand{\deletedeli}[2][]{%
	\ifx&#1&%
	\deleted[id=Eli]{#2}%
	\else
	\deleted[id=Eli, comment=#2]{#1}%
	\fi
}
\newcommand{\replacedeli}[3][]{%
	\ifx&#1&%
	\replaced[id=Eli]{#3}{#2}%
	\else
	\replaced[id=Eli, comment=#1]{#3}{#2}%
	\fi
}

\begin{document}
	\title{Quantum nonlocal correlations of anomalous weak values}
	\author{Ron Cohen}
	\affiliation{Faculty of Engineering and Institute of Nanotechnology and Advanced Materials, Bar-Ilan University, Ramat Gan 5290002, Israel}	
	\author{Avshalom C. Elitzur}	
	\affiliation{
		Institute for Quantum Studies, Chapman University, Orange, CA 92866, United
		States of America
	}
	\affiliation{
		Iyar, The Israeli Institute for Advanced Research, POB 651 Zichron Ya'akov
		3095303, Israel
	}
	\author{Eliahu Cohen}
	\affiliation{Faculty of Engineering and Institute of Nanotechnology and Advanced Materials, Bar-Ilan University, Ramat Gan 5290002, Israel}
	\affiliation{
		Institute for Quantum Studies, Chapman University, Orange, CA 92866, United
		States of America
	}
	\begin{abstract}
		Violations of Bell inequalities are a hallmark of entanglement, with only entangled states capable of exceeding classical bounds in standard Bell tests. Here we analyze anomalous weak values of the 
		Clauser--Horne--Shimony--Holt (CHSH) Bell operator in pre- and post-selected (PPS) quantum ensembles, using them to define separability-constrained bounds on Bell-type nonlocal correlations in the presence of post-selection. 
		Fixing the overlap between the pre- and post-selected states, we compare three scenarios: unrestricted boundary states, one separable boundary state, and both boundary states separable. For each case, we derive both the maximal weak value for a fixed Bell operator and the maximal bound obtained by further optimizing over all CHSH operators. Our results show that post-selection and entanglement are distinct operational resources: post-selection alone can enhance correlations, but entanglement is necessary to exceed the corresponding separable PPS bounds, and their combination yields the strongest attainable correlations. 
		Thus the separable PPS bound plays the role of a post-selected separability benchmark, distinct from the standard Bell bound.
		We further show that the enhancement beyond the separable bound closely tracks the concurrence of the states that optimize the bounds, identifying entanglement as the source of the additional correlation strength. 
		Finally, we show that nonlocal weak values provide post-selected entanglement witnesses, and we give a constructive protocol that detects every pure two-qubit source state with nonzero concurrence in the ideal state-adapted setting, even in regimes where the corresponding standard CHSH entanglement test is inconclusive.
		In this state-adapted setting, we explicitly construct the post-selection and CHSH measurements that achieve the largest possible separation from the separable PPS bound.
		More broadly, our results motivate hybrid protocols that combine post-selection and entanglement, with possible applications to improved quantum sensing, weak-value amplification, and quantum information processing.
	\end{abstract}
	
	\maketitle
	
	\section{Introduction}
	Quantum nonlocality is typically characterized by the violation of Bell inequalities---operational constraints obeyed by any locally causal model of spacelike-separated correlations \cite{Bell1964,BrunnerRMP2014}.
	Bell violations certify correlations incompatible with local hidden-variable theories while respecting no-signalling, and elevate entanglement from a kinematic property to a resource: all separable states satisfy every Bell inequality, whereas entanglement is necessary (though not sufficient) for a violation, which also requires incompatible measurements \cite{Werner1989,Horodecki1995}.
	Families of inequalities and loophole-free experiments,
	and	recent tests of proposed nonlocality limits 
	have mapped the quantum set---bounded by Tsirelson-type limits---between classical and broader no-signalling correlations \cite{Cirelson1980,PopescuRohrlich1994,Hensen2015,Giustina2015,Shalm2015,Navascues2007,Navascues2008,Teeni2019,atzori2024experimental}.
	From a device-independent perspective, Bell tests now serve as robust certification tools for states, measurements, randomness, and security \cite{MayersYao2004,Pironio2010,Acin2007}, a viewpoint we
	conceptually adopt for post-selected ensembles.
	However, post-selection alone can yield apparent Bell-inequality violations even for separable states (by conditioning on rare outcomes), seemingly undermining the special role of entanglement \cite{wharton2025bell,wang2025violation}. In this work we show that post-selection and entanglement are distinct resources, and that their combination strictly exceeds what post-selection alone can achieve.
	Within \emph{pre- and post-selected} (PPS) ensembles, weak values provide a natural probe of nonlocal correlations conditioned on both boundary states \cite{VaidmanTSVFReview2007,dressel2014colloquium}.
	This perspective has been used to analyze Einstein-Podolsky-Rosen-Bohm and additional scenarios of interest, where weak values demonstrate explicit nonlocal features \cite{AharonovCohen2015,aharonov2015can,elitzur2018nonlocal}. Moreover, post-selected ensembles can exhibit genuinely \emph{top-down} structure: higher-order correlations that cannot be reconstructed from lower-order marginals, highlighting a form of global nonlocal organization \cite{AharonovCohenTollaksenPNAS2018}. 
	Operationally, nonlocal observables are measurable in such settings \cite{VaidmanPRL2003}, and joint weak values of spatially separated observables can be extracted from correlations of local meter readouts \cite{ReschSteinbergPRL2004}--a procedure realized experimentally in Hardy-type settings and related photonic tests \cite{LundeenSteinbergPRL2009,CalderonLosadaCommPhys2020}.
	Related weak-measurement approaches have also been used to estimate Bell parameters while largely preserving the entanglement of the measured pair~\cite{white2016preserving,virzi2024entanglement}.
	Because post-selection alone can yield apparent violations in Bell-type tests, it may be tempting to conclude that entanglement is no longer essential in post-selected scenarios. Our standpoint is different: post-selection and entanglement are \emph{distinct} operational resources. Conditioning on successful post-selection reshapes the observed statistics of a pre- and post-selected ensemble, whereas entanglement endows the underlying bipartite system with genuinely nonclassical joint structure. We show that treating them on equal footing reveals a clear separation: when the ``amount of post-selection'' (e.g., the pre/post overlap) is held fixed, incorporating entanglement strictly enlarges the attainable correlations relative to what post-selection alone can produce. 
	This perspective clarifies that conditioning does not render entanglement superfluous; rather, entanglement and post-selection act as complementary resources whose joint use enables 
	higher
	correlations beyond what either can achieve alone.
	More broadly, this view is supported by earlier work on entanglement-assisted weak-value amplification, where entanglement enhances the efficiency of post-selected weak-value protocols in a different, metrological setting \cite{pang2014entanglement}.
	The interpretational status of weak measurements, in particular of anomalous weak values, has been debated, with claims that they merely reflect classical noise or selection bias rather than intrinsic quantum structure.
	Notably, anomalous weak values can be understood as operational proofs of contextuality under natural noncontextuality assumptions, reinforcing that their origin is genuinely nonclassical rather than a mere artefact of noise or post-selection \cite{pusey2014anomalous}.
	We address this by analyzing \emph{nonlocal} weak values of Bell-type observables in pre- and post-selected ensembles. 
	The appearance of correlated, setting-dependent weak values across spacelike-separated subsystems, together with their systematic dependence on entanglement versus separability, 
 	is constrained by the separability hierarchy derived below, and is therefore not captured merely by the possibility of post-selection-induced amplification.
	Accordingly, nonlocal weak values serve as operational probes of genuinely quantum correlations, reinforcing the physical significance of weak measurement beyond a ``measurement of noise'' narrative.
	In this work we quantify the respective roles of post-selection and entanglement by analyzing weak values of the Clauser--Horne--Shimony--Holt (CHSH) Bell operator in PPS ensembles at fixed pre/post overlap. First, for a fixed Bell operator, we derive overlap-constrained PPS bounds for three classes of boundary conditions: unrestricted pre- and post-selected states, one product boundary state, and two product boundary states. This yields a hierarchy of attainable transition amplitudes, and hence of weak values at fixed overlap, showing explicitly how separability constraints restrict the possible post-selected correlations. Second, we optimize these bounds over the full family of CHSH operators, obtaining the corresponding maximal PPS bounds and identifying the ultimate correlations achievable in each separability class.
	Third, we use the separable PPS bound as an operational entanglement-witness threshold. In particular, for pure two-qubit source states we identify a state-adapted protocol that detects every state with nonzero concurrence. We further prove in Appendix~\ref{Appendix:Global_optimality} that, when the product post-selection is freely optimized and each candidate is compared at its own overlap, it is sufficient and optimal to use a maximally incompatible CHSH operator: this choice maximizes the entanglement-detection margin above the separable PPS threshold.
	\\
	Beyond deriving the bounds, we use them to isolate the contribution of entanglement to PPS correlations. By comparing the unrestricted and one-product bounds with the two-product bound, we define the excess correlation enabled by entanglement and show that, for the optimizing families of states, this excess closely tracks their concurrence. Finally, we show that nonlocal weak values can be used as post-selected entanglement witnesses: violation of the two-product PPS bound certifies that at least one boundary state is entangled, while violation of the one-product bound can certify that both boundary states are entangled when one of them is independently characterized. 
	For pure source states, the resulting witness is constructive and state-adapted: the concurrence determines the optimal working overlap, and the optimal CHSH incompatibility can be chosen maximal.
	These results clarify how post-selection and entanglement combine in Bell-type PPS correlations, and suggest possible applications in weak-measurement-based certification, quantum sensing, and quantum information processing.
	The remainder of the paper is organized as follows.
	In Sec. \ref{Sec:CHSH_bounds_in_PPS_settings}
	we set the stage by introducing the two-state (pre- and post-selected) operator norms that underlie our analysis and by fixing the overlap parameter used for comparisons. 
	In Sec. \ref{Sec:CHSH_bounds_in_PPS_settings-General_Bounds}
	we derive general CHSH bounds for a fixed Bell operator (up to local unitaries), treating the cases of unrestricted, one-separable, and both-separable boundary states. In Sec. \ref{Sec:CHSH_bounds_in_PPS_settings-Maximal_Bounds} we obtain maximal bounds by jointly optimizing over both the boundary states and the Bell operator. 
	In Sec. \ref{Sec:Weak_values-Entanglement_Witness} we show that nonlocal weak values serve as entanglement witnesses.
	Section \ref{Sec:Conclusions} concludes with a summary of the results and an outlook.
	\section{CHSH bounds in PPS settings}
	\label{Sec:CHSH_bounds_in_PPS_settings}
\begin{figure}[h!]
		\begin{center}
			\includegraphics[scale=0.38]{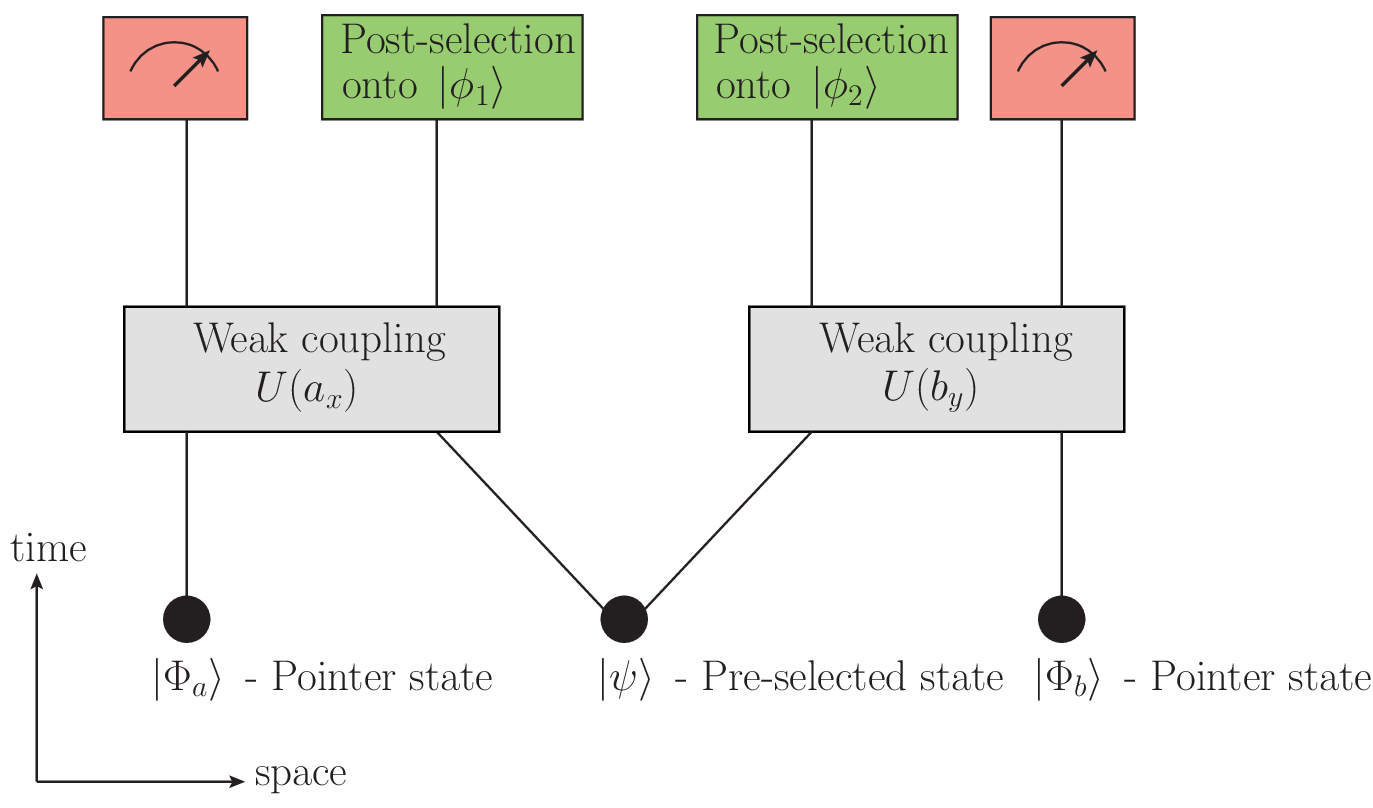}
			\caption{
				Schematic extraction of the joint weak value $(a_x\otimes b_y)_w$ with separable post-selection. 
				The bipartite system is pre-selected in $\ket{\psi}$, while Alice and Bob locally weakly couple their subsystems to pointer states $\ket{\Phi_a}$ and $\ket{\Phi_b}$ through interactions $U(a_x)$ and $U(b_y)$, respectively.
				After the weak interaction, local projective post-selections onto $\ket{\phi_1}$ and $\ket{\phi_2}$ impose the product final boundary state $\ket{\phi_1}\otimes\ket{\phi_2}$. 
				The joint weak value of $a_x\otimes b_y$ is reconstructed from the conditioned pointer data, including suitable two-pointer correlations and any single-pointer contributions required by the chosen weak-measurement scheme.
				Combining the reconstructed joint weak values for the relevant CHSH settings yields the weak value of the Bell operator ($B$).
			}
			\label{Fig:nonlocal_wv_protocol}
		\end{center}
	\end{figure}

	Here, we analyze the maximal weak values attainable in a CHSH scenario under various constraints, thereby extending the notion of quantum correlations to pre- and post-selected settings.
	Furthermore, we develop a two-state (pre- and post-selected) framework for quantifying quantum correlations in the presence of post-selection and for elucidating the role of entanglement in achieving the corresponding maximal bounds.
	We analyze three distinct scenarios.
	First, both the pre- and post-selected states are unrestricted, yielding a two-state analogue of the Tsirelson bound.
	Second, one of the states is constrained to be separable, and third, both states are restricted to be separable.
	We first recall that a general CHSH Bell operator has the form
	\begin{equation}
		B=a_0 \otimes (b_0+b_1) +a_1 \otimes (b_0-b_1)=\sum_{i,j=1}^{3}A_{ij}\sigma_i \otimes \sigma_j,	
	\end{equation}
	where $a_j=\vb{a}_j\cdot\vb*{\sigma}$ and $b_j=\vb{b}_j\cdot\vb*{\sigma}$ are dichotomic observables ($\pm1$ spectrum), $\vb*{\sigma}=(\sigma_1,\sigma_2,\sigma_3)$ denotes the Pauli vector in an arbitrary Cartesian basis, and $A_{ij}=\tr(\frac{B\sigma_i \otimes \sigma_j}{4})$.
	It can be shown by the singular-value decomposition of $A$, that 
	any CHSH operator is locally equivalent to the canonical form
	\begin{equation}\label{Eq:Local_Equivalent_B}
		B\equiv (U_a \otimes U_b)B(U_a \otimes U_b)^\dagger
		=
		\Lambda_+	\sigma_1
		\otimes\sigma_1
		+
		\Lambda_-	\sigma_2
		\otimes\sigma_2,	
	\end{equation}
	where $\Lambda_\pm
	=\sqrt{
		2\left(
		1\pm
		\sqrt{
			1-
			\gamma^2
		}
		\right)
	}
	$ are the singular values of $A$ and
	$\gamma=\norm{\vb{a}_0\cross \vb{a}_1}\norm{\vb{b}_0\cross \vb{b}_1},\,\,
	\gamma\in \left[0,1\right]$
	quantifies the local incompatibility of the measurement pairs.
	From an experimental perspective, the following measurement choices
	\begin{equation}
		\begin{aligned}
			\begin{aligned}
				\vb{a}_0&=(1,0,0),
				\\
				\vb{a}_1&=(0,1,0),
			\end{aligned}
			&&
			\begin{aligned}
				\vb{b}_0&=(\cos(\theta),\sin(\theta),0),
				\\
				\vb{b}_1&=(\cos(\theta),-\sin(\theta),0),
			\end{aligned}
		\end{aligned}	
	\end{equation} 
	where
	\begin{equation}
		\begin{aligned}
			&\cos(\theta)=\frac{\Lambda_+}{2}=\sqrt{\frac{1+\sqrt{1-\gamma^2}}{2}},
			\\&\sin(\theta)=\frac{\Lambda_-}{2}=\sqrt{\frac{1-\sqrt{1-\gamma^2}}{2}},
			&&\theta\in \left[0,\frac{\pi}{4}\right].
		\end{aligned}	
	\end{equation}
	can produce an equivalent form
	\begin{equation}\label{gen_bell1}
		\begin{aligned}
			&B=
			2\left(
			\cos(\theta)\sigma_1
			\otimes \sigma_1
			+
			\sin(\theta)\sigma_2
			\otimes \sigma_2
			\right).
		\end{aligned}	
	\end{equation}
	The form given by Eqs. (\ref{Eq:Local_Equivalent_B}) and (\ref{gen_bell1}) 
	is used throughout this analysis.
	The standard Tsirelson bound for the CHSH inequality can be evaluated by the operator norm of the Bell operator in question ($\opnorm{B}$). 
	We define the operator norm of a bounded operator $A$ acting on a Hilbert space $\mathcal{H}$ as
	\begin{equation}
		\opnorm{A}:=
		\sup
		\left\{
		\norm{A\ket{\psi}}:
		\norm{\ket{\psi}}=1, \ket{\psi}\in \mathcal{H}
		\right\}.
	\end{equation}
	Similarly, for a set of operators $\mathcal{A}$, we define $\opnorm{\mathcal{A}}$ as the supremum of the operator norm over $A\in \mathcal{A}$, namely
	\begin{equation}\label{Eq:Op_norm-on_set}
		\opnorm{\mathcal{A}}
		:=\sup
		\left\{
		\opnorm{A}: A\in \mathcal{A}
		\right\}.	
	\end{equation}
	The maximal Tsirelson bound can be derived from the supremum of the Tsirelson bound over the set of all Bell operators ($\mathcal{B}$), namely $\opnorm{\mathcal{B}}=2\sqrt{2}.$
This upper bound can only be attained using entangled states; if the state is restricted to be product, then the largest CHSH expectation value obtainable after optimizing over all CHSH operators is 2, coinciding with the usual classical CHSH bound.
	Similarly, by post-selecting the final state, weak values of a Bell operator can be measured.
	\\\\
	The weak value of an operator $A$ with respect to pre-selected state $\ket{\psi}$ and post-selected state $\ket{\phi}$ is defined as
	\begin{equation}
		A_w=\frac{\bra{\phi}A\ket{\psi}}{\braket{\phi}{\psi}}.	
	\end{equation}
	The magnitude $\abs{A_w}=\frac{\abs{\bra{\phi}A\ket{\psi}}}{c}$, depends on the overlap between the pre- and post-selected states $c:=\abs{\braket{\phi}{\psi}} \in (0,1]$.
	Consequently, $\abs{A_w}$ can essentially be arbitrarily large by having a very small overlap.
	Therefore, in order to compare fairly between the cases we consider here,
	we assume that the magnitude of the overlap remains constant across all cases, i.e., $c=\abs{\braket{\phi}{\psi}}=\abs{\braket{\phi_s}{\psi}} = \abs{\braket{\phi_s}{\psi_s}}$, 
where the subscript indicates a product-state constraint, and we ask which boundary-state class yields the larger weak value.
	\\
	With this convention, comparing weak values at a fixed overlap $c$ reduces to comparing the corresponding transition amplitudes $\abs{\bra{\phi}A\ket{\psi}}$. The associated weak-value bounds are then recovered by dividing the transition-amplitude bounds by $c$. This separates the nontrivial effect of the separability constraints from the trivial amplification obtained by taking the overlap arbitrarily small.
	\\\\
	In the CHSH setting considered below, the relevant operator $A$ will be the Bell operator $B$, which is a linear combination of product observables $a_x\otimes b_y$. Operationally, $B_w$ can therefore be reconstructed from the joint weak values $(a_x\otimes b_y)_w$ associated with the different CHSH settings. Each such joint weak value can be reconstructed using only local weak couplings to $a_x$ and $b_y$, followed by conditioning on the chosen final boundary state. A schematic realization for the case of separable post-selection is shown in Fig.~\ref{Fig:nonlocal_wv_protocol}.
	\\\\
	We can therefore define new types of two-state operator norm
	\begin{widetext} 
		\begin{equation}
			\begin{aligned}
				\twonorm{A}{c}
				&:=
				\sup
				\left\{
				\abs{
					\bra{\phi}
					A
					\ket{\psi}
				}: \norm{\ket{\phi}}=\norm{\ket{\psi}}=1,\,\,\abs{\braket{\phi}{\psi}}=c,\,\,\,
				\left\{\ket{\psi},\ket{\phi}\right\}\in \mathcal{H}
				\right\},
				\\
				\twonorm{A}{c,1s}
				&:=
				\sup
				\left\{
				\abs{
					\bra{\phi_s}
					A
					\ket{\psi}
				}: \norm{\ket{\phi_s}}=\norm{\ket{\psi}}=1,\,\,\abs{\braket{\phi_s}{\psi}}=c,\,\,
				\ket{\phi_s}\,\,\text{separable}
				,\,\,\,
				\left\{\ket{\psi},\ket{\phi_s}\right\}\in \mathcal{H}
				\right\},
				\\
				\twonorm{A}{c,s}
				&:=
				\sup
				\left\{
				\abs{
					\bra{\phi_s}
					A
					\ket{\psi_s}
				}: \norm{\ket{\phi_s}}=\norm{\ket{\psi_s}}=1,\,\,\abs{\braket{\phi_s}{\psi_s}}=c,\,\,
				\left\{\ket{\psi_s},\ket{\phi_s}\right\}\,\,\text{separable}
				,\,\,\,
				\left\{\ket{\psi_s},\ket{\phi_s}\right\}\in \mathcal{H}
				\right\}
				.
			\end{aligned}
		\end{equation}
	\end{widetext}
The natural two-state extension of the operator norm is $\twonorm{A}{c}$, which plays the role of a Tsirelson-type transition-amplitude bound for PPS correlations at fixed overlap. We also define the half-separable bound $\twonorm{A}{c,1s}$, where one boundary state is constrained to be product, and the fully separable bound $\twonorm{A}{c,s}$, where both boundary states are constrained to be product.
	\\
	It is important to note that
	\begin{equation}
		\twonorm{A}{c,s}
		\leq
		\twonorm{A}{c,1s}
		\leq
		\twonorm{A}{c}
		\leq \opnorm{A}.	
	\end{equation}
The corresponding unrestricted weak-value bound for operator $A$ at fixed overlap $c$ is therefore
	\begin{equation}
		\norm{A}_{w,c}:=\frac{
			\twonorm{A}{c}
		}{c}.
	\end{equation}
	For each of these two-state operator norms, we 
	define their actions on a set of operators $\mathcal{A}$ in the same fashion as in Eq. (\ref{Eq:Op_norm-on_set}), namely
	\begin{equation}\label{Eq:TS-op_norm-on_set}
		\begin{aligned}
			\twonorm{\mathcal{A}}{c}
			&:=\sup
			\left\{
			\twonorm{A}{c} : A\in \mathcal{A}
			\right\},
			\\
			\twonorm{\mathcal{A}}{c,1s}
			&:=\sup
			\left\{
			\twonorm{A}{c,1s} : A\in \mathcal{A}
			\right\},
			\\
			\twonorm{\mathcal{A}}{c,s}
			&:=\sup
			\left\{
			\twonorm{A}{c,s} : A\in \mathcal{A}
			\right\}.
		\end{aligned}
	\end{equation}
	Consequently, the maximal two-state bounds (obtained by optimizing over the set of Bell operators $\mathcal{B}$) are 
	$\twonorm{\mathcal{B}}{c}$,
	$\twonorm{\mathcal{B}}{c,1s}$, and
	$\twonorm{\mathcal{B}}{c,s}$.
	In the following, we analyze these bounds.
	As a consistency check, the analytic expressions derived below were also verified by independent numerical optimizations over the corresponding state and measurement parameters.
	Before proceeding, we emphasize that the present analysis is restricted to pure pre- and post-selected boundary states. 
	The generalization of these bounds to mixed PPS states is left for future work.
	\subsection{General bounds}
	\label{Sec:CHSH_bounds_in_PPS_settings-General_Bounds}
	Here, we analyze the $\twonorm{B}{c},\,\twonorm{B}{c,1s}$ and $\twonorm{B}{c,s}$, for the Bell operator given by Eq. (\ref{Eq:Local_Equivalent_B})
	and an arbitrary value of $\gamma$.
	In addition, we consider the entangled states which set these bounds and show a correspondence between their entanglement measure (concurrence) and their difference with respect to the 
	separable 
	bound $\twonorm{B}{c,s}$.
	This correspondence highlights the role of entanglement in the enhancement of post-selected correlations.
	\subsubsection{Tsirelson bound}
	Here we derive the two-state Tsirelson bound $\twonorm{B}{c}$ for an arbitrary Bell operator.
	We start by rewriting the Bell operator in the following way
	\begin{equation}
		\begin{aligned}
			B=
			&\left(\Lambda_++\Lambda_-\right)
			\frac{
				\sigma_1\otimes\sigma_1+\sigma_2\otimes\sigma_2
			}{2}
			\\	+
			&	\left(\Lambda_+-\Lambda_-\right)
			\frac{
				\sigma_1\otimes\sigma_1-\sigma_2\otimes\sigma_2
			}{2}.
		\end{aligned}	
	\end{equation}
	We can identify the following
	\begin{equation}
		\begin{aligned}
			\frac{
				\sigma_1\otimes\sigma_1+\sigma_2\otimes\sigma_2
			}{2}
			&=
			\ketbra{\Phi_{1+}}{\Phi_{1+}}-\ketbra{\Phi_{1-}}{\Phi_{1-}},
			\\
			\frac{
				\sigma_1\otimes\sigma_1-\sigma_2\otimes\sigma_2
			}{2}
			&=
			\ketbra{\Phi_{2+}}{\Phi_{2+}}-\ketbra{\Phi_{2-}}{\Phi_{2-}},
		\end{aligned}	
	\end{equation}
	where the set $\left\{\Phi_{1\pm},\Phi_{2\pm}\right\}$ is an orthonormal basis.
	To see this is true in general \cite{chefles1997diagonalisation}, as an example, we can pick $\sigma_1=X,\,\sigma_2=Y$ and get
	\begin{equation}
		\begin{aligned}
			\frac{
				X\otimes X+Y\otimes Y
			}{2}
			&=
			\ketbra{\psi^+}{\psi^+}-\ketbra{\psi^-}{\psi^-},
			\\
			\frac{
				X\otimes X-Y\otimes Y
			}{2}
			&=
			\ketbra{\phi^+}{\phi^+}-\ketbra{\phi^-}{\phi^-},
		\end{aligned}	
	\end{equation}
	where $\left\{\psi^\pm,\phi^\pm\right\}$ is the set of Bell states.
	The Bell operator now reads
	\begin{equation}
		\begin{aligned}
			B=&\lambda_+
			\left(
			\ketbra{\Phi_{1+}}{\Phi_{1+}}-\ketbra{\Phi_{1-}}{\Phi_{1-}}
			\right)	
			\\	+
			&\lambda_-
			\left(
			\ketbra{\Phi_{2+}}{\Phi_{2+}}-\ketbra{\Phi_{2-}}{\Phi_{2-}}
			\right),
		\end{aligned}
	\end{equation}
	where we denoted $\lambda_\pm=\Lambda_+\pm\Lambda_-=2\sqrt{1\pm\gamma}$.
	The parameter $\gamma$ is directly related to the operator norm, since $\opnorm{B}=\lambda_+$.
	Earlier we stated that the two-state operator norm, is bounded by the standard operator norm ($\twonorm{B}{c}\leq\opnorm{B}$).
	Here we will show that this bound can always be saturated for any value of the overlap $c$, namely
	\begin{equation}
		\twonorm{B}{c}=\opnorm{B}=\lambda_+.
	\end{equation}
	To see that, we consider the following states
	\begin{equation}\label{Eq:Tsirelson_states}
		\begin{aligned}
			&
			\ket{\psi}=
			\frac{1}{\sqrt{2}}
			\left(
			\sqrt{1+c}\ket{\Phi_{1+}}+\sqrt{1-c}\ket{\Phi_{1-}}
			\right),
			\\&\ket{\phi}=
			\frac{1}{\sqrt{2}}
			\left(
			\sqrt{1+c}\ket{\Phi_{1+}}-\sqrt{1-c}\ket{\Phi_{1-}}
			\right),
		\end{aligned}	
	\end{equation}
	for which
	\begin{equation}
		\begin{aligned}
			&
			\braket{\phi}{\psi}=c,
			&&
			\bra{\phi}
			B
			\ket{\psi}=\lambda_+.
		\end{aligned}
	\end{equation}
	Therefore
	\begin{equation}
		\label{Eq:fixed_tsirelson_bound}
		\begin{aligned}
			&\twonorm{B}{c}=\lambda_+=\opnorm{B},
			&&
			\norm{B}_{w,c}=
			\frac{\lambda_+}{c}.
		\end{aligned}	
	\end{equation}
	\subsubsection{Half-separable case}
	Here we derive the two-state operator norm $\twonorm{B}{c,1s}$ with the constraint that one of the states is separable.
	We assume that the post-selected state $\ket{\phi_s}=\ket{\phi_1}\otimes \ket{\phi_2}$ is a product state, i.e. can be obtained via a local projective measurement of each particle.
	\\
	We begin by constructing the pre-selected state as
	\begin{equation}
		\ket{\psi}=c\ket{\phi_s}
		+\sqrt{1-c^2}\ket{\phi_{s,\perp} (\rho)},
	\end{equation}
	where $\ket{\phi_{s,\perp} (\rho)}$ represent any (normalized) state which is orthogonal to $\ket{\phi_s}$ and $\rho$ is the set of free parameters defining such state, i.e. 
	any state from the set of normalized orthogonal (to $\ket{\phi_s}$) states can be achieved by an appropriate choice of parameters $\rho$.
	Using this construction we get
	\begin{equation}\label{Eq:Half-separable}
		\bra{\phi_s}B\ket{\psi}=c
		\bra{\phi_s}
		B
		\ket{\phi_s}
		+\sqrt{1-c^2}	\bra{\phi_s}B\ket{\phi_{s,\perp} (\rho)}.
	\end{equation}
	We can bound the second term, by applying the Aharonov-Vaidman identity \cite{aharonov1990properties,leifer2023uncertainty}
	\begin{equation}
		A\ket{\Psi}
		=	
		\expval{A}_{\Psi}\ket{\Psi}
		+
		\Delta\left(A\right)_{\Psi}\ket{\Psi_{s,\perp}^{(A)}},	
	\end{equation}
	where
	\begin{equation}
		\begin{aligned}
			\expval{A}_{\Psi}=
			\bra{\Psi}
			A
			\ket{\Psi},
			\,\,
			\Delta\left(A\right)_{\Psi}=
			\sqrt{
				\bra{\Psi}
				A^\dagger A
				\ket{\Psi}
				-
				\bra{\Psi}
				A
				\ket{\Psi}^2
			}
			.
		\end{aligned}	
	\end{equation}
	and get
	\begin{equation}
		\begin{aligned}
			\bra{\phi_s}B\ket{\phi_{s,\perp} (\rho)}
			&=
			\left(
			\expval{B}_{\phi_s}\bra{\phi_s}
			+
			\Delta\left(B\right)_{\phi_s}\bra{\phi_{s,\perp}^{(B)}}
			\right)\ket{\phi_{s,\perp} (\rho)}
			\\&=
			\Delta\left(B\right)_{\phi_s}
			\braket{\phi_{s,\perp}^{(B)}}{\phi_{s,\perp} (\rho)},
		\end{aligned}
	\end{equation}
	Therefore, we find
	\begin{equation}
		\abs{
			\bra{\phi_s}B\ket{\phi_{s,\perp} (\rho)}	
		}
		\leq
		\Delta\left(B\right)_{\phi_s}	.
	\end{equation}
	Since we are free to choose $\ket{\phi_{s,\perp} (\rho)}$ to be any state which is orthonormal to $\ket{\phi_s}$ and in particular $\ket{\phi_{s,\perp} (\rho)}=\ket{\phi_{s,\perp}^{(B)}}=
	\frac{
		B-\expval{B}_{\phi_s}
	}{\Delta\left(B\right)_{\phi_s}}\ket{\phi_s}
	$, 
	this bound can always be saturated, and we obtain the following general relation
	\begin{equation}\label{gen-separable-relation}
		\abs{\bra{\phi_s}B\ket{\psi}}
		\leq
		c \abs{\expval{B}_{\phi_s}}+\sqrt{1-c^2}\Delta\left(B\right)_{\phi_s}.	
	\end{equation}
	Next, we use Eq. (\ref{gen_bell1}) to rewrite the Bell operator as
	\begin{equation}\label{gen_bell2}
		\begin{aligned}
			&B=
			2\left(
			\cos(\theta)X
			\otimes X
			+
			\sin(\theta)Y
			\otimes Y
			\right),
			\\	&
			B^2=4\left(
			1-\sin(2\theta) Z \otimes Z
			\right),
		\end{aligned}	
	\end{equation}
	where we picked $(\sigma_1,\sigma_2,\sigma_3)=(X,Y,Z)$.
	Using the relation given in Eq. (\ref{gen-separable-relation}) and following an optimization procedure (see Appendix \ref{Appendix:half-separable}), we obtain the ``half-separable'' bound
	\begin{equation}\label{Eq:HSG}
		\twonorm{B}{c,1s}=
		\begin{cases}
			\lambda_+\sqrt{1-c^2} & 0\leq c \leq c_0\\
			\sqrt{\lambda_+^2+\Lambda_-^2}-\Lambda_-c
			& c_0\leq c \leq c_1 \\
			\Lambda_+c+\Lambda_-\sqrt{1-c^2} & c_1\leq c \leq 1
		\end{cases},	
	\end{equation}	
	where 
	$
	c_0=\sqrt{\frac{\Lambda_-^2}{
			\lambda_+^2+\Lambda_-^2	
	}}
	$ and
	$
	c_1=\sqrt{\frac{\lambda_+^2}{
			\lambda_+^2+\Lambda_-^2	
	}}
	$.
	Illustration of this result can be found in Appendix \ref{Appendix:Gen-Bounds} (see Fig. \ref{Fig:two-state-HSG}).
	\subsubsection{Separable case}
	Here we derive the two-state operator norm $\twonorm{B}{c,s}$ with the constraint that both pre- and post-selected states are separable, namely
	\begin{equation}
		\ket{\phi}=\ket{\phi_1}\otimes\ket{\phi_2}
		,
		\ket{\psi}=\ket{\psi_1}\otimes\ket{\psi_2}.	
	\end{equation}
	The overlap between the states is given by
	\begin{equation}
		c=
		\abs{\braket{\phi}{\psi}}=
		\abs{\braket{\phi_1}{\psi_1}
			\braket{\phi_2}{\psi_2}}=r_1r_2,	
	\end{equation}
	where we denote $\abs{\braket{\phi_j}{\psi_j}}=r_j$ and demand that $r_1 r_2=c$.
	To proceed, we use Eq. (\ref{gen_bell2}) and get
	\begin{equation}
		\label{Eq:optimize-separable}
		\begin{aligned}
			\bra{\phi}
			B
			\ket{\psi}
			&=
			2\left[\cos(\theta)
			X_1
			X_2
			+
			\sin(\theta)
			Y_1
			Y_2
			\right]
			,
		\end{aligned}	
	\end{equation}
	where
	\begin{equation}
		\begin{aligned}
			X_j=\bra{\phi_j}X\ket{\psi_j},
			\,\,\,\,
			Y_j=\bra{\phi_j}Y\ket{\psi_j},
			\,\,\,\,
			Z_j=\bra{\phi_j}Z\ket{\psi_j}.
		\end{aligned}	
	\end{equation}
	Following an optimization procedure (see Appendix \ref{Appendix:separable}), we obtain the separable bound
	\begin{equation}
		\label{Eq:SG}
		\begin{aligned}
			\twonorm{B}{c,s}
			&=
			2\left[\cos(\theta)
			+
			\sin(\theta)
			(1-c)
			\right]
			\\&	=
			\lambda_+-\Lambda_-c
			.
		\end{aligned}	
	\end{equation}
	The two-state norm $\twonorm{B}{c,s}$ under the constraint of separability is linearly decreasing from 
	$\lambda_+$ ($c=0$) to $\Lambda_+$ ($c=1$).
	Illustration of this result can be found in Appendix \ref{Appendix:Gen-Bounds} (see Fig. \ref{Fig:two-state-SG}).
	Incidentally, these values at the boundary ($c=0,1$) coincide with the values given by $\twonorm{B}{c,1s}$, namely
	\begin{equation}
		\begin{aligned}
			&\twonorm{B}{c=0,s}=\twonorm{B}{c=0,1s}=\lambda_+,
			\\&\twonorm{B}{c=1,s}=\twonorm{B}{c=1,1s}=\Lambda_+.
		\end{aligned}	
	\end{equation}
	\subsubsection{Summary}
\begin{figure}[h!]
		\begin{center}
			\includegraphics[scale=0.75]{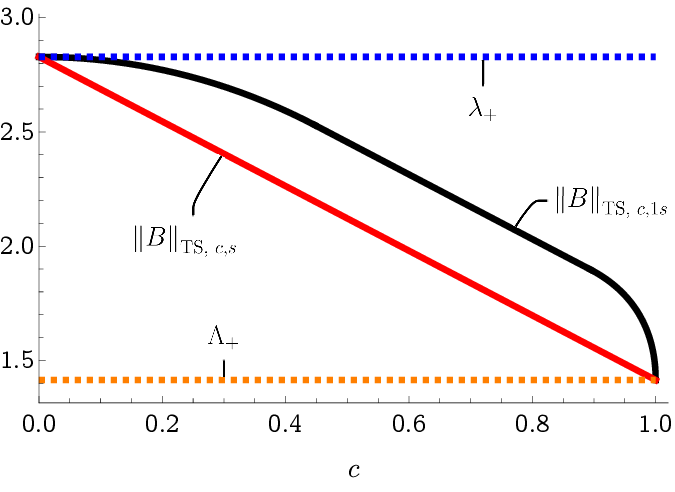}
			\caption{
				Two-state operator norm for the Bell operator $B_{\gamma=1}$ under the separability constraint for one (black solid line) and two states (red solid line) as a function of the overlap $c$.
				In the dashed blue line is the two-state Tsirelson bound for $\gamma=1$ without any constraints which is equivalent to the standard operator norm.
				In dashed orange, is the 
				standard separable bound,
				namely the maximal value of 
				$
				\bra{\phi_s}
				B_{\gamma=1}
				\ket{\phi_s}
				=\sqrt{2}$.	
			}
			\label{Fig:two-state-G}
		\end{center}
	\end{figure}

	We compare the three cases for a representative Bell operator with $\gamma=1$ in Fig. \ref{Fig:two-state-G}.
	We see a clear hierarchy between the three cases.
	When there is no entanglement (red curve), the correlation strength increases linearly as the overlap decreases, illustrating a trade-off between correlation strength and the probability to post-select.
	On the other hand, when either (black curve) or both (dashed blue) of the PPS states are entangled, the correlations surpass what is achievable without entanglement.
	This suggests that the contribution coming strictly from entanglement is hidden in the difference between the curves, namely
	\begin{equation}\label{Eq:Entanglement_contribution}
		\begin{aligned}
			\delta\twonorm{B}{c}&:=\twonorm{B}{c}-\twonorm{B}{c,s},
			\\
			\delta\twonorm{B}{c,1s}&:=\twonorm{B}{c,1s}-\twonorm{B}{c,s}.
		\end{aligned}
	\end{equation}
	Interestingly, these differences closely track the entanglement strength required to generate the corresponding correlations.
	More concretely, we compare them to the concurrence generated by the entangled state(s) in each case.
	\begin{equation}\label{Eq:Concurrence}
		\begin{aligned}
			&\mathcal{C}_1(\ket{\psi})=\mathcal{C}_1(\ket{\phi})=c,
			\\&
			\mathcal{C}_2(\ket{\psi})=
			\begin{cases}
				2c_0 c_1
				& c_0\leq c \leq c_1 \\
				2c\sqrt{1-c^2} & \text{else}
			\end{cases},
		\end{aligned}	
	\end{equation}
	where $\mathcal{C}_1$ 
	is the concurrence of
	either of states in Eq. (\ref{Eq:Tsirelson_states}), $\mathcal{C}_2$ 
	is the concurrence of a
	representative state $\ket{\psi}$ which optimizes $\twonorm{B}{c,1s}$, and $(c_0,c_1)$ are as defined following Eq. (\ref{Eq:HSG}).
	\begin{figure}[h!]
		\begin{center}
			\includegraphics[scale=0.45]{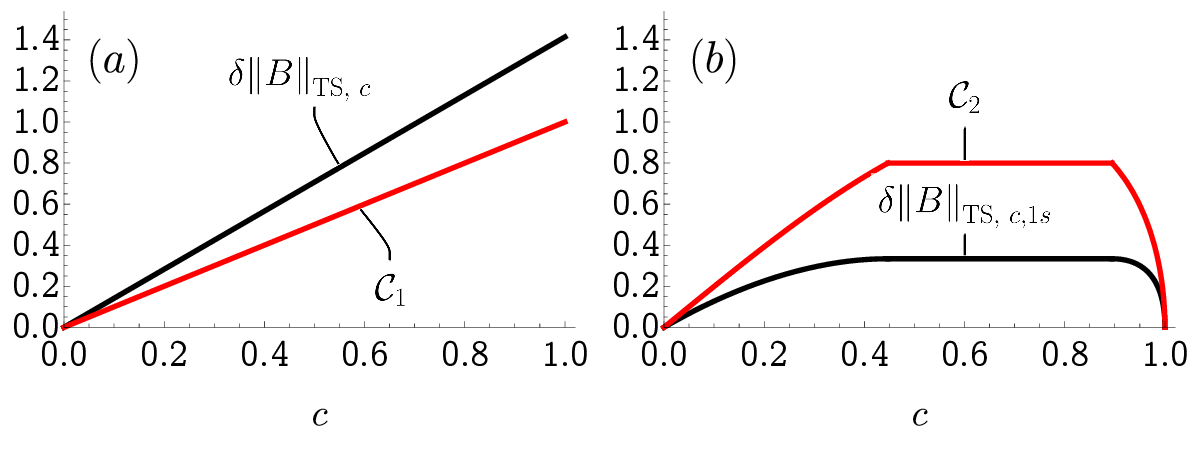}
			\caption{
				Correspondence between degree of entanglement 
				and the 
				contribution of entanglement to the two-state bounds 
				$\twonorm{B}{c}$ and $\twonorm{B}{c,1s}$, for $\gamma=1$ as a function of the overlap.
				(a)
				Unrestricted two-state (Tsirelson) case: the concurrence $\mathcal{C}_1$
				of either boundary state (red) is compared with 
				$\delta\twonorm{B}{c}$ 
				(black).
				(b) Half-separable case: the concurrence $\mathcal{C}_2$ of a representative $\ket{\psi}$ --- the entangled state that optimizes $\twonorm{B}{c,1s}$ (red) ---
				is compared with 
				$\delta\twonorm{B}{c,1s}$ (black).
				The close tracking of these pairs highlights how the additional correlations beyond the
				separable PPS bound
				are
				tied to the degree of entanglement in the PPS boundary states.
			}
			\label{Fig:Correlation-entanglement-correspondence}
		\end{center}
	\end{figure}
	The correspondence between the degree of entanglement (concurrence) of the states that set $\left(\twonorm{B}{c},\twonorm{B}{c,1s}\right)$ and their difference from the separable
	bound $\twonorm{B}{c,s}$, defined in
	Eq. (\ref{Eq:Entanglement_contribution}),
	is depicted in Fig. \ref{Fig:Correlation-entanglement-correspondence}, for 
	$\gamma=1$.
	The two graphs, clearly show qualitative similarities between $\left(\delta\twonorm{B}{c},\delta\twonorm{B}{c,1s}\right)$ and their corresponding concurrences $\left(\mathcal{C}_1,\mathcal{C}_2\right)$, indicating that these indeed capture the contribution of entanglement to the bounds.
	In the Tsirelson case, the ability to convert entanglement into correlations 
	is optimal, since $\twonorm{B}{c}=\opnorm{B}$ for every overlap $c$.
	When the overlap is minimal, most of the correlation comes from post-selection alone (without entanglement), since in the limit of zero overlap even the 
	separable
	bound saturates $\opnorm{B}$. Entanglement cannot be used to enhance it further
	because the correlations are physically capped by $\opnorm{B}$. 
	In the opposite limit of almost no post-selection ($c\rightarrow1$), entanglement is mostly responsible for reaching the bound.
	In the half-separable case, the story is more involved.
	On the one hand, entanglement is necessary to enhance the correlation beyond the 
	separable
	bound, but it does so less efficiently than in the Tsirelson case.
	As seen in Fig. \ref{Fig:Correlation-entanglement-correspondence} (a), the enhancement $\delta\twonorm{B}{c}$ (black curve) lies above the corresponding concurrence $\mathcal{C}_1$ (red curve), whereas in Fig. \ref{Fig:Correlation-entanglement-correspondence} (b) the opposite is true
	for $\delta\twonorm{B}{c,1s}$ and $\mathcal{C}_2$. This indicates that the amount of correlation obtained from a given amount of entanglement is lower than what is possible in the unrestricted case. The shortfall stems from the separability constraint imposed on one of the boundary states.
	\subsection{Maximal bounds}
	\label{Sec:CHSH_bounds_in_PPS_settings-Maximal_Bounds}
	In the previous section, we derived the fixed-operator bounds $\twonorm{B}{c}$, $\twonorm{B}{c,1s}$, and $\twonorm{B}{c,s}$. We now optimize these quantities over the CHSH family $\mathcal{B}$, obtaining the maximal bounds $\twonorm{\mathcal{B}}{c}$, $\twonorm{\mathcal{B}}{c,1s}$, and $\twonorm{\mathcal{B}}{c,s}$.
	\subsubsection{Tsirelson bound}
For the unrestricted two-state bound, Eq.~(\ref{Eq:fixed_tsirelson_bound}) gives $\twonorm{B}{c}=\lambda_+=2\sqrt{1+\gamma}=\opnorm{B}$. Maximizing over CHSH operators therefore amounts to maximizing $\gamma$, which is attained at $\gamma=1$. Hence
	\begin{equation}
		\twonorm{\mathcal{B}}{c}=\opnorm{\mathcal{B}}=2\sqrt{2}.	
	\end{equation}
	\subsubsection{Half-separable case}
To derive $\twonorm{\mathcal{B}}{c,1s}$ we optimize the three branches of Eq.~(\ref{Eq:HSG}) over $\theta$, or equivalently over $k=\tan\theta\in[0,1]$. For a fixed value of $k$, the branch boundaries are
\begin{equation}\label{Eq:max_half_sep_boundaries}
	c_0(k)=\frac{k}{\sqrt{1+2k+2k^2}},
	\qquad
	c_1(k)=\frac{1+k}{\sqrt{1+2k+2k^2}} .
\end{equation}
The first branch is proportional to $\lambda_+=2\sqrt{1+\gamma}$ and is therefore maximized at $\gamma=1$, i.e. $k=1$. This branch is available only when $c\leq c_0(1)=1/\sqrt{5}$, giving
\begin{equation}\label{Eq:max_half_sep_first_branch}
	\twonorm{\mathcal{B}}{c,1s}=2\sqrt{2}\sqrt{1-c^2},
	\qquad 0\leq c\leq \frac{1}{\sqrt{5}}.
\end{equation}
For $c>1/\sqrt{5}$ the first branch is no longer available for any CHSH operator, so the optimum must lie either on the second branch or on the third branch. The third branch satisfies
\begin{equation}\label{Eq:max_half_sep_third_branch_bound}
	2\left(c\cos\theta+\sqrt{1-c^2}\sin\theta\right)
	\leq 2
\end{equation}
by Cauchy--Schwarz. Thus it is enough to maximize the second branch, provided that the optimized second branch is at least $2$. Writing the second branch as
\begin{equation}\label{Eq:max_half_sep_H2}
	H_2(k,c)=
	\frac{2\left(\sqrt{1+2k+2k^2}-kc\right)}{\sqrt{1+k^2}},
\end{equation}
we find
\begin{equation}\label{Eq:max_half_sep_stationary}
	\frac{\partial H_2(k,c)}{\partial k}=0
	\quad\Longleftrightarrow\quad
	c\sqrt{1+2k+2k^2}=1+k-k^2 .
\end{equation}

For $c\in[1/\sqrt{5},1]$, this equation has a unique solution $\kappa(c)\in[0,1]$, with $\kappa(1/\sqrt{5})=1$ and $\kappa(1)=0$.
Moreover, this solution satisfies
$c_0(\kappa(c))\leq c\leq c_1(\kappa(c))$, so the second branch is
self-consistent; see Appendix~\ref{App:MaxHalfSepAux}. 
Since $H_2(0,c)=2$, the optimized second branch obeys $H_2(\kappa(c),c)\geq2$ and therefore dominates the third branch.
Hence we conclude
	\begin{equation}\label{Eq:HSM}
		\twonorm{\mathcal{B}}{c,1s}=
		\begin{cases}
			2\sqrt{2}\sqrt{1-c^2} & 0\leq c \leq \frac{1}{\sqrt{5}}\\
			H_2(\kappa(c),c)
			& \frac{1}{\sqrt{5}}\leq c\leq 1 
		\end{cases}.	
	\end{equation}
	The two-state maximal bound $\twonorm{\mathcal{B}}{c,1s}$ is depicted by the black curve in Fig. \ref{Fig:two-state-M}.
	\subsubsection{Separable case}
\begin{figure}[h!]
		\begin{center}
			\includegraphics[scale=0.75]{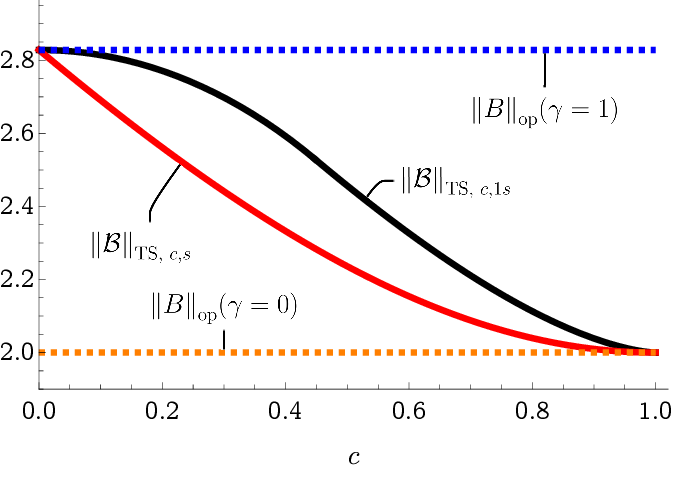}
			\caption{
				Two-state maximal operator norm (optimized for all CHSH Bell operators) under the separability constraint for one (black solid line) and two states (red solid line) as a function of the overlap $c$.
				In the dashed blue line is the two-state maximal Tsirelson bound which is equivalent to the standard maximal Tsirelson bound.
				In dashed orange, is the 
				maximal standard separable bound,
				namely
				$
				\sup\limits_{B\in\mathcal B}\beta_{\text{sep}}(B)=2
				$.
			}
			\label{Fig:two-state-M}
		\end{center}
	\end{figure}

	To derive $\twonorm{\mathcal{B}}{c,s}$ we use our result from Eq. (\ref{Eq:SG}) and optimize over all possible Bell operators, which is equivalent to optimizing over $\theta$.
	For this purpose, we set $u=\cos(\theta)$ and find solution for
	\begin{equation}
		\begin{aligned}
			&\pdv{\twonorm{B}{c,s}}{u}=0,
			&&\Rightarrow u=\frac{1}{\sqrt{1+(1-c)^{2}}}.	
		\end{aligned}
	\end{equation}
	Plugging back the solution we find 
	\begin{equation}
		\twonorm{\mathcal{B}}{c,s}
		=2\sqrt{1+(1-c)^2}.	
	\end{equation}
	The two-state maximal bound $\twonorm{\mathcal{B}}{c,s}$ is depicted by the red curve in Fig. \ref{Fig:two-state-M}.
	\section{Nonlocal weak values as entanglement witnesses}
	\label{Sec:Weak_values-Entanglement_Witness}
	In the previous section we showed that entanglement is essential for achieving stronger post-selected correlations than are possible with separable 
	boundary states. While post-selection alone can increase the measured correlations, supplementing it with entanglement boosts them even further 
	(see Figs. \ref{Fig:two-state-G} and \ref{Fig:two-state-M}).
	This implies that weak values of a Bell observable can provide evidence for the presence of entanglement in either or both the pre- and post-selected states.
	%
	For comparison, consider the fixed-setting CHSH entanglement witness
	obtained from the same Bell operator $B$. Let $\mathrm{Prod}(\mathcal H)$ denote the set of normalized pure product states in the bipartite Hilbert space $\mathcal H$.
	The corresponding separable expectation value bound is
	\begin{equation}\label{Eq:standard_separable_bound}
		\beta_{\text{sep}}(B)
		:=\sup_{\ket{\phi_s}\in
			\text{Prod}(\mathcal{H})	
		} \abs{\bra{\phi_s}B\ket{\phi_s}}=\Lambda_+.	
	\end{equation}
	This is the so-called ``\textit{standard separable bound}"  which certifies entanglement for a state $\ket{\psi}$
	if its corresponding CHSH expectation value exceeds it, namely
	$\abs{\bra{\psi}B\ket{\psi}}>\beta_{\text{sep}}(B)$.
	Entanglement certification using weak values must rely on a different bound, since
	any weak value can formally exceed the \textit{standard} separable bound (Eq. \ref{Eq:standard_separable_bound})
	simply by having a sufficiently small overlap.
	Here entanglement detection is enabled by exceeding the	\textit{post-selected separable bound}---the maximal weak value attainable when both boundary states are separable. 
	\\
	
	We now make this intuition precise. Fix a Bell operator $B$ and an overlap $c=\abs{\braket{\phi}{\psi}}$ between pre- and post-selected states.
	From Sec. \ref{Sec:CHSH_bounds_in_PPS_settings} we know that, if both boundary states are separable, the transition amplitude $\abs{\bra{\phi_s}B\ket{\psi_s}}$ is bounded by
	\begin{equation}
		\abs{\bra{\phi_s}B\ket{\psi_s}}\leq	\twonorm{B}{c,s}.
	\end{equation}
	Consequently, the corresponding weak value obeys
	\begin{equation}\label{Eq:two_state-separable_inequality}
		\abs{B_w}=\frac{\abs{\bra{\phi_s}B\ket{\psi_s}}}{\abs{\braket{\phi_s}{\psi_s}}}
		\leq \frac{\twonorm{B}{c,s}}{c}.
	\end{equation}
	Any experimentally observed weak value that violates 
	Eq. (\ref{Eq:two_state-separable_inequality}), i.e.,
	obeys the following inequality
	\begin{equation}\label{Eq:detection_criteria}
		\abs{B_w}>\frac{\twonorm{B}{c,s}}{c},
	\end{equation}
	cannot arise from any pair of separable pre- and post-selected states with the given overlap. Therefore, such a weak value serves as a \textit{post-selected entanglement witness}: at least one of $\ket{\psi}$ or $\ket{\phi}$ must be entangled.\\
	Moreover, Eq. (\ref{Eq:detection_criteria}) can be made state-specific when one of the boundary states is independently characterized. If one boundary state is known to be separable, then a violation of Eq. (\ref{Eq:detection_criteria}) certifies that the other boundary state is entangled. Conversely, if one boundary state is known to be entangled, Eq. (\ref{Eq:detection_criteria}) alone does not identify whether the other boundary state is entangled as well. In that case one should compare the observed weak value with the half-separable threshold. Since $\twonorm{B}{c,1s}$ is the maximal transition amplitude attainable when one boundary state is separable and the other is unrestricted, any weak value satisfying
	\begin{equation}\label{Eq:detection_criteria_2}
		\abs{B_w}>\frac{\twonorm{B}{c,1s}}{c}
	\end{equation}
	rules out all PPS descriptions in which either boundary state is separable. Thus, given an independently characterized entangled post-selection, Eq. (\ref{Eq:detection_criteria_2}) certifies that the pre-selected state is entangled as well. This condition is more demanding than Eq. (\ref{Eq:detection_criteria}), since \(\twonorm{B}{c,s}\leq\twonorm{B}{c,1s}\), but it certifies the stronger statement that one entangled boundary state is not sufficient to explain the observed weak value.
	\\\\
	In an experimental implementation, the witness can be used either in a targeted or in an adaptive mode. If a model for the prepared source state $\ket{\psi}$ is available, one may choose in advance the Bell operator ($B$) and the product post-selection ($\ket{\phi_s}$) expected to maximize the witness margin
	\begin{equation}\label{Eq:witness_margin}
		\margin (B,\phi_s)
		=c\abs{B_w}-\twonorm{B}{c,s}.	
	\end{equation}
	For a fixed choice of $B$ and $\ket{\phi_s}$, entanglement is certified when the measured margin is positive with sufficient statistical significance. If no reliable model is available, one may first perform an exploratory search over CHSH settings and product post-selections to identify a promising candidate, and then repeat the experiment with this candidate fixed in order to estimate $B_w$, $c$, and hence $\margin$ with the required accuracy. Equivalently, one may correct the final significance for the number of candidates tested. 
	In either case, failure to violate the threshold should be interpreted as inconclusive rather than as evidence of separability, since it may simply reflect a nonoptimal choice of Bell operator, post-selection, or overlap, insufficient signal-to-noise ratio, or other experimental imperfections.
	It is important to note that, within the pure-state setting considered here, the same reasoning applies after exchanging the roles of the pre- and post-selected boundary states. Since $B$ is Hermitian, both $\abs{\bra{\phi}B\ket{\psi}}$ and $\abs{\braket{\phi}{\psi}}$ are invariant under this exchange. Thus, if the pre-selected state is known and separable, violation of Eq.~(\ref{Eq:detection_criteria}) certifies entanglement of the post-selected boundary state. More generally, if the known pre-selected state is entangled, then certifying entanglement of the post-selected state requires comparison with the half-separable threshold in Eq.~(\ref{Eq:detection_criteria_2}).
	\subsubsection*{Constructive $\gamma=1$ witness for pure states}
\begin{figure}[t]
		\centering
		\includegraphics[width=1\linewidth]{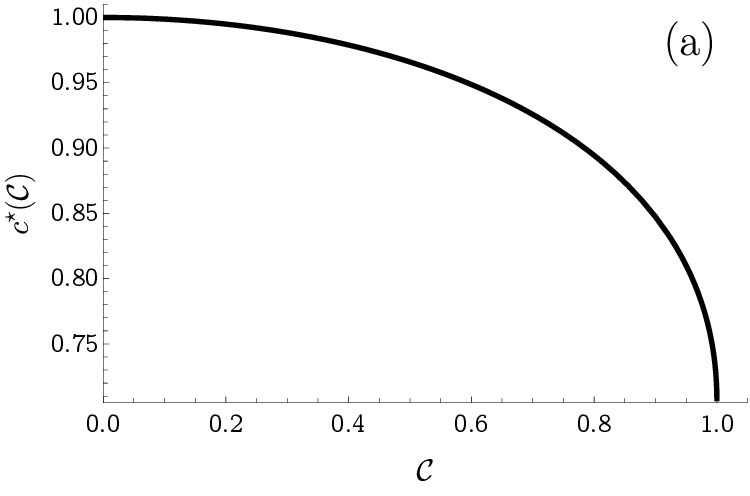}
		\includegraphics[width=1\linewidth]{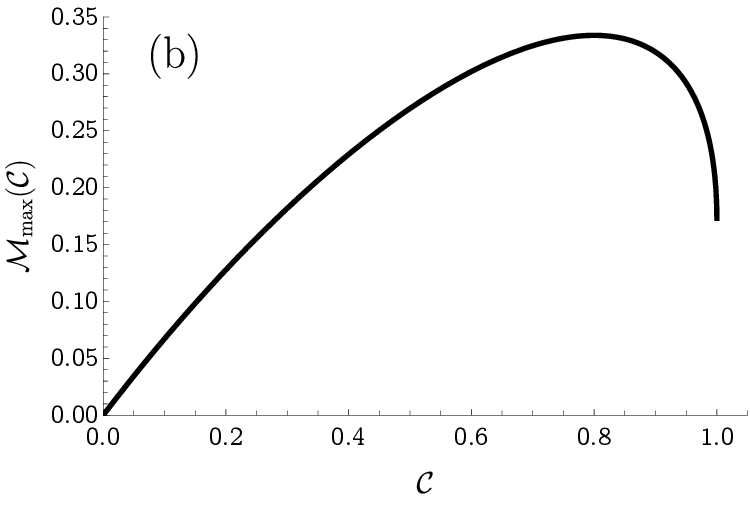}
		\caption{
			State-adapted design parameters for the constructive $\gamma=1$ weak-value witness.
			(a) Optimal working overlap $c^\star(\mathcal C)$ as a function of the concurrence $\mathcal C$.
			Weakly entangled states require high-overlap post-selection, $c^\star\to1$ as $\mathcal C\to0$, while maximally entangled states correspond to $c^\star=1/\sqrt2$.
			(b) Corresponding maximal transition-amplitude witness margin $\mathcal M_{\max}(\mathcal C)$.
			The margin is strictly positive for every $\mathcal C>0$, showing that every pure entangled two-qubit source state is detectable in principle. The maximal margin occurs at $\mathcal C=4/5$, with $\mathcal M_{\max}=\sqrt2(\sqrt5-2)$.
		}
		\label{Fig:cstar_margin_concurrence}
	\end{figure}

	Having established the general witness criteria, we now show that they lead to a constructive detection strategy for arbitrary pure entangled two-qubit states. In particular, we show that, up to local unitary rotations, it is sufficient to use a maximally incompatible CHSH operator with $\gamma=1$ together with a suitable product post-selection. We first justify the choice $\gamma=1$ within the Schmidt-aligned construction and then show that the resulting witness detects every pure entangled two-qubit state.
	Consider an arbitrary pure entangled two-qubit state written in its Schmidt basis as
	\begin{equation}
		\label{Eq:Schmidt_state_witness}
		\ket{\psi}=\alpha\ket{00}+\beta\ket{11},\qquad
		\alpha\geq\beta>0,\qquad \alpha^2+\beta^2=1.
	\end{equation}
	Choose the product post-selection $\ket{\phi_s}=\ket{00}$, so that $c=\abs{\braket{\phi_s}{\psi}}=\alpha$. For the Schmidt-aligned CHSH family
	\begin{equation}
		\label{Eq:Schmidt_aligned_CHSH}
		B_{\theta}=2\left(\cos(\theta)Z\otimes Z+\sin(\theta)X\otimes X\right),\qquad \theta\in\left[0,\frac{\pi}{4}\right],
	\end{equation}
	where $\gamma=\sin(2\theta)$, the transition amplitude is
	\begin{equation}
		\abs{\bra{\phi_s}B_{\theta}\ket{\psi}}
		=2\left(\alpha\cos(\theta)+\beta\sin(\theta)\right).
	\end{equation}
	On the other hand, the corresponding separable PPS bound is
	\begin{equation}
		\twonorm{B_{\theta}}{c,s}=2\left[\cos(\theta)+\sin(\theta)(1-\alpha)\right].
	\end{equation}
	Therefore the transition-amplitude witness margin for this construction is
	\begin{equation}
		\begin{aligned}
			\margin_{\theta}
			&=\abs{\bra{\phi_s}B_{\theta}\ket{\psi}}-\twonorm{B_{\theta}}{c,s}
			\\&=2\left[(\alpha-1)\cos(\theta)+(\alpha+\beta-1)\sin(\theta)\right].
		\end{aligned}
	\end{equation}
	Its derivative with respect to $\theta$ is
	\begin{equation}
		\frac{d\margin_{\theta}}{d\theta}
		=2\left[(1-\alpha)\sin(\theta)+(\alpha+\beta-1)\cos(\theta)\right].
	\end{equation}
	For every entangled state, $\beta>0$ implies $\alpha<1$ and $\alpha+\beta>1$, while $\sin(\theta),\cos(\theta)\geq0$ on $\theta\in[0,\pi/4]$. Hence $d\margin_{\theta}/d\theta>0$ throughout this interval, and the margin is maximized at $\theta=\pi/4$, equivalently $\gamma=1$. 
	This establishes optimality within the Schmidt-aligned construction. In Appendix~\ref{Appendix:Global_optimality} we prove the stronger state-adapted statement: for every fixed $\theta$, no local CHSH frame and no product post-selection can exceed this margin when each candidate is compared at its corresponding overlap $c=\abs{\braket{\phi_s}{\psi}}$. Consequently, when the overlap is not externally fixed but is chosen through the product post-selection, the maximal transition-amplitude witness margin is globally attained at $\gamma=1$.
	At this optimal point,
	\begin{equation}
		B_{\gamma=1}=\sqrt{2}\left(Z\otimes Z+X\otimes X\right),
	\end{equation}
	and the transition amplitude becomes
	\begin{equation}
		\abs{\bra{\phi_s}B_{\gamma=1}\ket{\psi}}
		=\sqrt{2}(\alpha+\beta).
	\end{equation}
	The separable PPS bound at the same overlap is
	\begin{equation}
		\twonorm{B_{\gamma=1}}{c,s}=\sqrt{2}(2-c)=\sqrt{2}(2-\alpha).
	\end{equation}
	Consequently, the witness condition is equivalent to
	\begin{equation}
		\sqrt{2}(\alpha+\beta)>\sqrt{2}(2-\alpha),
	\end{equation}
	or
	\begin{equation}
		2\alpha+\beta>2.
	\end{equation}
	This inequality is strict for every $\beta>0$, since
	\begin{equation}
		2\alpha+\beta-2=\beta\left(1-\frac{2\beta}{1+\alpha}\right)>0,
	\end{equation}
	where we used $\alpha\geq\beta$ and $\beta<1$ to obtain $1+\alpha>2\beta$. Therefore every pure entangled two-qubit state can, in principle, be detected by a $\gamma=1$ CHSH weak-value witness with a suitable product post-selection. The corresponding working overlap is
	\begin{equation}
		\label{Eq:constructive_working_overlap}
		c^{\star}=\alpha=\sqrt{\frac{1+\sqrt{1-\mathcal C^2}}{2}},
	\end{equation}
	where $\mathcal C=2\alpha\beta$ is the concurrence. Thus weakly entangled states are detected near high-overlap post-selections, $c^{\star}\rightarrow1$, whereas a maximally entangled state corresponds to $c^{\star}=1/\sqrt{2}$.
	The corresponding maximal state-adapted transition-amplitude margin can be written as
	\begin{equation}
		\label{Eq:Mmax_concurrence}
		\margin_{\max}(\mathcal C)
		=
		\sqrt{2}
		\left[
		2\sqrt{\frac{1+\sqrt{1-\mathcal C^2}}{2}}
		+
		\sqrt{\frac{1-\sqrt{1-\mathcal C^2}}{2}}
		-
		2
		\right].
	\end{equation}
	Fig.~\ref{Fig:cstar_margin_concurrence} displays both the optimal working overlap and this maximal margin as design parameters for the state-adapted witness.
	\subsubsection*{Adaptive implementation}	
	The preceding construction also suggests how the general margin in Eq. (\ref{Eq:witness_margin}) can be optimized in practice. The statistical significance mentioned above can be quantified as follows. If $\hat{\margin}$ is the experimentally estimated margin and $\sigma_{\margin}$ is its uncertainty, then a one-sided confidence criterion may be expressed as
	\begin{equation}
		\hat{\margin}-z_{\eta}\sigma_{\margin}>0,
	\end{equation}
	or equivalently $\hat{\margin}/\sigma_{\margin}>z_{\eta}$, 
	where $\eta$ is the chosen one-sided significance level and $z_{\eta}$ is the corresponding threshold for the statistical distribution of the estimator.
	For example, under an approximately Gaussian error model, $z_{\eta}=3$ corresponds to a three-sigma one-sided criterion. The uncertainty $\sigma_{\margin}$ should include the statistical uncertainty in the reconstructed weak value and in the overlap $c$, as well as relevant systematic uncertainties. In the $\gamma=1$ implementation, Eq. (\ref{Eq:witness_margin}) reduces to
	\begin{equation}
		\margin(B_{\gamma=1},\phi_s)=c\abs{B_w}-\sqrt{2}(2-c).
	\end{equation}
	
	Here the choice of $B$ means the choice of local CHSH measurement directions, or equivalently the local CHSH frame; this is independent of the choice of the product final boundary state $\ket{\phi_s}=\ket{\phi_A}\otimes\ket{\phi_B}$. We denote by $\sigma_{\vb{n}}:=\vb{n}\cdot\vb*{\sigma}$ the Pauli observable along the Bloch-sphere direction $\vb{n}$. For $\gamma=1$, a local CHSH frame can be specified by choosing two orthogonal Bloch-sphere directions $\vb{u}_{A},\vb{v}_{A}$ for Alice and two orthogonal Bloch-sphere directions $\vb{u}_{B},\vb{v}_{B}$ for Bob, giving the locally equivalent Bell operator
	\begin{equation}
		B_{\gamma=1}=\sqrt{2}\left(
		\sigma_{\vb{u}_{A}}\otimes\sigma_{\vb{u}_{B}}
		+
		\sigma_{\vb{v}_{A}}\otimes\sigma_{\vb{v}_{B}}
		\right).
	\end{equation}
	Equivalently, this corresponds to the CHSH settings $a_0=\sigma_{\vb{u}_{A}}$, $a_1=\sigma_{\vb{v}_{A}}$, $b_0=(\sigma_{\vb{u}_{B}}+\sigma_{\vb{v}_{B}})/\sqrt{2}$, and $b_1=(\sigma_{\vb{u}_{B}}-\sigma_{\vb{v}_{B}})/\sqrt{2}$, up to local unitary rotations.
	
	If a reliable model for the source is available, the constructive proof above suggests a natural starting point. For a state approximately written in Schmidt form as $\ket{\psi}=\alpha\ket{00}+\beta\ket{11}$, with $\alpha\geq\beta$, one may start from the product post-selection $\ket{\phi_s}=\ket{00}$ and the Schmidt-aligned Bell operator $B_{\gamma=1}=\sqrt{2}(Z\otimes Z+X\otimes X)$. This choice is motivated by the fact that $\ket{00}$ is the dominant product component, while $X\otimes X$ connects it to the correlated component $\ket{11}$ responsible for the entanglement contribution to the transition amplitude.
	The global optimality result of Appendix~\ref{Appendix:Global_optimality} further justifies fixing $\gamma=1$ for the ideal pure-state, state-adapted optimization in which the overlap is selected by the product post-selection.
	If no reliable model is available, one may use an adaptive search. First, fix $\gamma=1$ and scan over a finite set of local CHSH frames and product post-selections. The scan need not be restricted to post-selections with a fixed overlap: each candidate $\ket{\phi_s}$ determines its own overlap $c=\abs{\braket{\phi_s}{\psi}}$ in the ideal pure-state setting, and is compared with the corresponding separable threshold at that value of $c$. For each candidate pair $(B,\ket{\phi_s})$, one estimates the post-selection probability, reconstructs $B_w$ from the locally measured joint weak values, and computes the corresponding margin $\margin(B,\phi_s)$ or the statistically normalized margin $\margin/\sigma_{\margin}$. A practical coarse search may begin with several local single-qubit Pauli frames: on Alice's and Bob's sides, one can choose the two orthogonal CHSH directions from the local axis pairs $\{Z,X\}$, $\{Z,Y\}$, and $\{X,Y\}$. These choices can be combined with product post-selections formed from the corresponding local Pauli eigenstates. If preliminary local tomography or marginal information is available, the search can instead be seeded by the dominant local eigenstates of the reduced states.
	The best candidates can then be refined over nearby local measurement directions and nearby product states on the two local Bloch spheres. Once a promising candidate has been identified, the final witness test should either be performed with that candidate fixed on an independent set of runs, or the reported significance should be corrected for the number of candidates tested. 
	\subsection{Post-selected detection vs. standard detection}
	In Sec. \ref{Sec:CHSH_bounds_in_PPS_settings} we identified two families of entangled boundary states that saturate the two-state bounds for a fixed Bell operator $B$ with $\gamma>0$. 
	The first family, given in Eq. (\ref{Eq:Tsirelson_states}), saturates the two-state Tsirelson bound and is defined for $0<c\leq 1$. 
	The second family is associated with the entangled states that saturate the half-separable bound $\twonorm{B}{c,1s}$ and is defined for $0<c<1$. 
	In a PPS experiment, both families are detectable by their corresponding nonlocal weak values, as expressed in Eq. (\ref{Eq:detection_criteria}). 
	However, if one performs a CHSH test for the purpose of entanglement detection
	using the \emph{same} measurement settings (i.e., the same operator $B(\gamma)$) and evaluates the usual CHSH expectation value $\mathcal{S}(\ket{\psi}):=\abs{\bra{\psi}B\ket{\psi}}$, 
	then a violation of  
	$\mathcal{S}(\ket{\psi})\leq \beta_{\text{sep}}(B)$ 
	(see Eq.~(\ref{Eq:standard_separable_bound})) implies that $\ket{\psi}$ is entangled; nevertheless, only a subset of these states achieves such a violation for the fixed settings.
	\begin{figure}[h!]
		\begin{center}
			\includegraphics[scale=0.75]{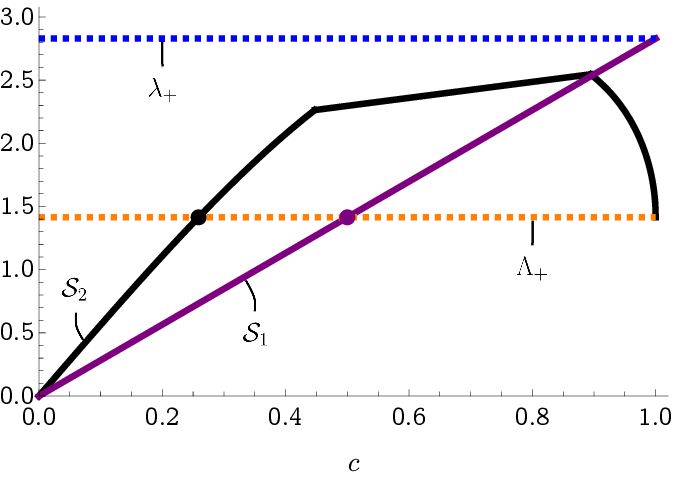}
			\caption{
				Standard CHSH test values for the two families of entangled states that maximize the PPS bounds for a fixed Bell operator $B_{\gamma=1}$. The dashed blue line is the Tsirelson/operator-norm limit $\lambda_{+}=\|B\|_{\mathrm{op}}=2\sqrt{2}$. The dashed orange line is the standard 
				separable 
				limit for this fixed operator, $\beta_{\text{sep}}=\Lambda_+=\sqrt{2}$. The solid purple curve shows $\mathcal{S}_1(c)$ (see Eq. (\ref{Eq:S_1})), obtained from either of the Tsirelson-saturating boundary states (18), and the solid black curve shows $\mathcal{S}_2(c)$ (see Eq. (\ref{Eq:S_2})) for a representative entangled state that saturates the half-separable bound. The highlighted markers indicate the corresponding detection thresholds for a standard CHSH test: $\mathcal{S}_1$ 
				allows detection only
				for 
				$c>c_{\text{threshold}}^{(1)}=\frac{1}{2}$
				(purple point), whereas $\mathcal{S}_2$ 
				allows detection only
				for $c_{\text{threshold}}^{(2)}<c<1$ with $c_{\text{threshold}}^{(2)}\approx0.259$ (black point). Thus, although both families are entangled throughout their domains, standard detection succeeds only in restricted overlap ranges.
			}
			\label{Fig:standard-detection}
		\end{center}
	\end{figure}
	This distinction is illustrated in Fig. \ref{Fig:standard-detection} for $\gamma=1$, where we plot the standard CHSH values $\mathcal{S}_1(c)$ and $\mathcal{S}_2(c)$ associated with the Tsirelson-saturating family and the half-separable--saturating family, respectively (see Appendix \ref{Appendix:standard_detection}). 
	Although both families are entangled throughout their domains, 
	a standard CHSH test certifies this entanglement only in restricted overlap ranges:
	$\mathcal{S}_1(c)$ 
	exceeds the standard separable bound $\beta_{\text{sep}}(B)$
	only for $c_{\text{threshold}}^{(1)}<c\leq 1$, whereas $\mathcal{S}_2(c)$ violates it only for $c_{\text{threshold}}^{(2)}<c<1$.
	In contrast, the post-selected weak-value criterion detects these families across their full domains. 
	Therefore, even for fixed measurement settings and for the specific entangled states that optimize the PPS bounds, post-selection can render entanglement operationally detectable in regimes where the corresponding standard CHSH test remains inconclusive.
	\section{Conclusions}
	\label{Sec:Conclusions}
	We have developed a separability-constrained two-state framework for Bell-type correlations in pre- and post-selected quantum ensembles. Within this framework, anomalous weak values of the CHSH Bell operator define a hierarchy of PPS bounds, depending on whether none, one, or both boundary states are required to be separable. For each case we derived both fixed-setting bounds and maximal bounds optimized over all CHSH operators. The resulting structure shows unambiguously that entanglement remains essential in post-selected Bell scenarios: while post-selection alone can generate large weak values, it is entanglement that enables one to exceed the separable two-state bounds. We further identified the entanglement contribution by comparing the unrestricted and half-separable bounds with the fully separable one, and found that this excess closely follows the concurrence of the optimizing states.
	In addition, we showed that nonlocal weak values furnish post-selected entanglement witnesses, since violating the separable two-state weak-value threshold certifies entanglement in at least one boundary state.
	For pure two-qubit source states, we further provided a constructive protocol showing that every state with nonzero concurrence can be detected in principle by a suitable product post-selection. 
	Moreover, in the state-adapted setting where the product post-selection is optimized freely, the construction is globally optimal over local CHSH frames and product post-selections, and the maximal margin is attained with a maximally incompatible CHSH operator.
	The corresponding witness margin closes continuously in the product limit, highlighting that weakly entangled states pose a finite-statistics challenge rather than a conceptual limitation of the witness.
	The central operational message is that anomalous nonlocal weak values are not merely large conditional averages: at fixed overlap, they obey a separability hierarchy that can be used as an entanglement witness.
	The significance of these results is twofold. First, they clarify a conceptual issue that has become increasingly important in light of recent discussions of Bell-type violations generated by post-selection. Post-selection can reshape observed statistics dramatically, but it does not make entanglement superfluous. Rather, once the amount of post-selection is fixed through the overlap, post-selection and entanglement emerge as distinct operational ingredients, with entanglement supplying the extra correlation strength beyond what separable PPS states can achieve. Second, the witness perspective shows that weak values are not merely a byproduct of conditioning, but can serve as an operational probe of entanglement in nonlocal settings. In particular, the PPS witness can detect entanglement for families of states and fixed measurement settings for which the corresponding standard Bell test does not yet yield a conclusive violation.
	These observations open several avenues for future work. A first direction is to generalize the present separability-constrained analysis to other Bell inequalities, higher-dimensional systems, and multipartite PPS settings. A second is to investigate robustness against noise, imperfect overlap estimation, and finite statistics, thereby moving from an ideal witness criterion to an experimentally deployable one. A third concerns applications: the present results suggest that post-selection and entanglement can be combined in hybrid weak-measurement protocols, potentially linking Bell-type PPS correlations with metrological and sensing advantages. More broadly, our analysis supports the view that anomalous nonlocal weak values provide a useful bridge between foundational questions about quantum correlations and operational tasks based on pre- and post-selected ensembles.
	
	\section*{Acknowledgments} We thank Ken Wharton for helpful comments. This work was supported by the European Union's Horizon Europe research and innovation programme under grant agreement No. 101178170, by the Israel Science Foundation under grant agreement No. 2208/24 and by the Israeli Council for Higher Education through ``QERNEL'' and the ``Interdisciplinary Center for the Theory of Quantum Computing''.
	\appendix
	\section{General bounds}
	\label{Appendix:Gen-Bounds}
	\begin{figure}[h!]
		\begin{center}
			\includegraphics[scale=0.65]{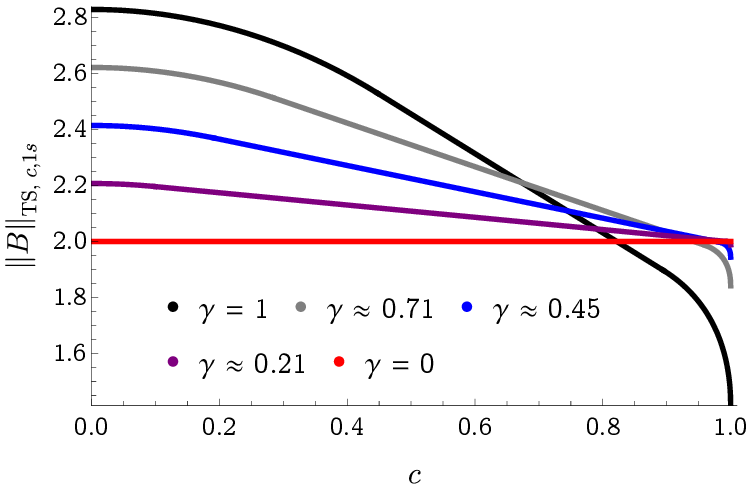}
			\caption{
				Two-state operator norm for a general Bell operator $B(\gamma)$ under the separability constraint for one state
				as a function of the overlap $c$.
			}
			\label{Fig:two-state-HSG}
		\end{center}
	\end{figure}
	\begin{figure}[h!]
		\begin{center}
			\includegraphics[scale=0.65]{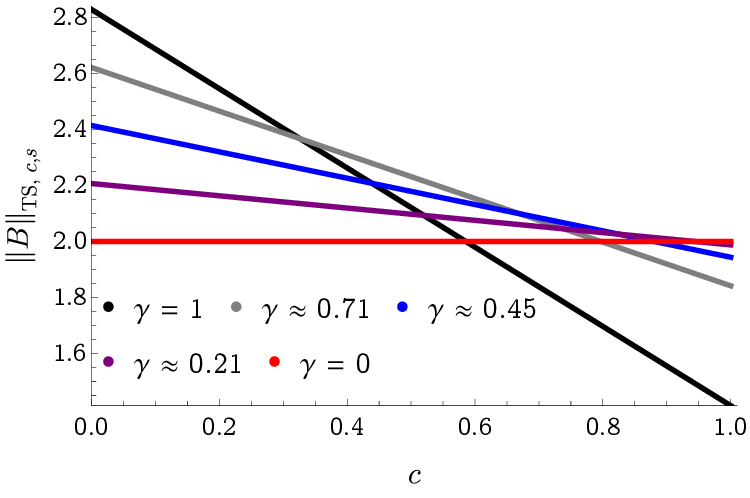}
			\caption{
				Two-state operator norm for a general Bell operator $B(\gamma)$ under the separability constraint for both pre- and post-selected states
				as a function of the overlap $c$.
			}
			\label{Fig:two-state-SG}
		\end{center}
	\end{figure}
	\section{Optimization for the ``half-separable'' case}
	\label{Appendix:half-separable}
	The bound for the ``half-separable'' case can be found by optimizing
	\begin{equation}\label{Eq:AppB_H}
		H=c \abs{\expval{B}_{\phi_s}}+\sqrt{1-c^2}\Delta\left(B\right)_{\phi_s},	
	\end{equation}
	over all product post-selections $\ket{\phi_s}=\ket{\phi_1}\otimes\ket{\phi_2}$.
	Let
	\begin{equation}
		a=\cos\theta,
		\quad
		b=\sin\theta,
		\quad
		0\leq b\leq a,
	\end{equation}
	and write
	\begin{equation}
		X_j=\bra{\phi_j}X\ket{\phi_j},
		\,
		Y_j=\bra{\phi_j}Y\ket{\phi_j},
		\,
		Z_j=\bra{\phi_j}Z\ket{\phi_j}.
	\end{equation}
Since $\ket{\phi_j}$ is a single-qubit pure state, $(X_j,Y_j,Z_j)$ is a unit Bloch vector. Using Eq. (\ref{gen_bell2}),
	\begin{equation}\label{Eq:AppB_expectation_variance}
	\begin{aligned}
		\expval{B}_{\phi_s}
		&=
		2\left(aX_1X_2+bY_1Y_2\right),
		\\
		\Delta(B)_{\phi_s}
		&=
		2\sqrt{
			1-2abZ_1Z_2-
			\left(aX_1X_2+bY_1Y_2\right)^2
		}.
	\end{aligned}
\end{equation}
Define
\begin{equation}
	T=\abs{aX_1X_2+bY_1Y_2},
	\qquad
	p=\abs{X_1X_2}+\abs{Y_1Y_2}.
\end{equation}
Since $a\geq b$, every product post-selection satisfies
\begin{equation}\label{Eq:AppB_T_bound}
	T
	\leq
	a\abs{X_1X_2}+b\abs{Y_1Y_2}
	\leq
	a\left(\abs{X_1X_2}+\abs{Y_1Y_2}\right)
	=
	ap.
\end{equation}
Thus, any configuration producing a given value of $T$ must satisfy $p\geq T/a$.
On the other hand, by the Cauchy-Schwarz inequality, and using the fact that each local post-selected state has a unit Bloch vector, we have
\begin{equation}\label{Eq:AppB_CS}
	\abs{X_1X_2}+\abs{Y_1Y_2}+\abs{Z_1Z_2}
	\leq
	1.
\end{equation}
Therefore
\begin{equation}
	\abs{Z_1Z_2}
	\leq
	1-p
	\leq
	1-\frac{T}{a}.
\end{equation}
For fixed $T$, the variance term in Eq. (\ref{Eq:AppB_expectation_variance}) is maximized by making $Z_1Z_2$ as negative as possible. Hence
\begin{equation}\label{Eq:AppB_variance_bound}
	1-2abZ_1Z_2-T^2
	\leq
	1+2ab\left(1-\frac{T}{a}\right)-T^2.
\end{equation}
Since $0\leq T\leq a$, we may write $T=ax$ with $x\in[0,1]$. Combining Eqs. (\ref{Eq:AppB_H}) and (\ref{Eq:AppB_variance_bound}) gives the upper bound
\begin{equation}\label{Eq:AppB_one_parameter_bound}
	H
	\leq
	2\left(
	cax
	+
	\sqrt{1-c^2}
	\sqrt{1+2ab(1-x)-a^2x^2}
	\right).
\end{equation}
This upper bound is tight for every $x\in[0,1]$. Indeed, it is saturated by the product states whose local Bloch vectors satisfy
\begin{equation}\label{Eq:AppB_saturating_bloch}
	Y_1=Y_2=0,
	\qquad
	X_1=X_2=\sqrt{x},
	\qquad
	Z_1=-Z_2=\sqrt{1-x}.
\end{equation}
For this family, $T=ax$ and $Z_1Z_2=-(1-x)$. Thus the optimization over all product post-selections reduces rigorously to the one-parameter maximization of
\begin{equation}\label{Eq:AppB_H_one_parameter}
	\small
	H(x)
	=
	2\left(
	c\cos\theta\,x
	+
	\sqrt{1-c^2}
	\sqrt{1+\sin(2\theta)(1-x)-\cos[2](\theta)x^2}
	\right).
\end{equation}
Taking the derivative with respect to $x$ gives the stationary points
\begin{equation}\label{Eq:AppB_xpm}
	\begin{aligned}
		\pdv{H}{x}=0
		\quad\Rightarrow\quad
		x_\pm
		&=
		\frac{
			\pm c\sqrt{1+\sin(2\theta)+\sin[2](\theta)}-
			\sin\theta
		}{\cos\theta}
		\\
		&=
		\pm c
		\sqrt{\frac{(\Lambda_++\Lambda_-)^2+\Lambda_-^2}{\Lambda_+^2}}
		-
		\frac{\Lambda_-}{\Lambda_+}.
	\end{aligned}
\end{equation}
Only $x_+$ can maximize $H(x)$ in the interval $0\leq x\leq1$. The boundary values at which $x_+=0$ and $x_+=1$ are
\begin{equation}\label{Eq:c-coefficients}
	\begin{aligned}
		c_0
		&=
		\sqrt{\frac{\sin[2](\theta)}{1+\sin(2\theta)+\sin[2](\theta)}}
		=
		\sqrt{\frac{\Lambda_-^2}{\lambda_+^2+\Lambda_-^2}},
		\\
		c_1
		&=
		\sqrt{\frac{1+\sin(2\theta)}{1+\sin(2\theta)+\sin[2](\theta)}}
		=
		\sqrt{\frac{\lambda_+^2}{\lambda_+^2+\Lambda_-^2}},
	\end{aligned}
\end{equation}
where $(\Lambda_++\Lambda_-)^2=\lambda_+^2$. Hence the maximizing value of $x$ is
\begin{equation}\label{Eq:x}
	x=
	\begin{cases}
		0   & 0\leq c \leq c_0,\\
		x_+ & c_0\leq c \leq c_1,\\
		1 & c_1\leq c \leq 1.
	\end{cases}
\end{equation}
Substituting these three branches into Eq. (\ref{Eq:AppB_H_one_parameter}) yields
	\begin{widetext} 
		\begin{equation}\label{Eq:HSG2}
			\twonorm{B}{c,1s}=
			\begin{cases}
				2\sqrt{1+\sin(2\theta)}\sqrt{1-c^2}=\lambda_+\sqrt{1-c^2} & 0\leq c \leq c_0\\
				2\left(\sqrt{1+\sin(2\theta)+\sin[2](\theta)}-\sin(\theta)c\right)=
				\sqrt{\lambda_+^2+\Lambda_-^2}-\Lambda_-c
				& c_0\leq c \leq c_1 \\
				2\left(c\cos(\theta)+\sqrt{1-c^{2}}\sin(\theta)\right)
				=\Lambda_+c+\Lambda_-\sqrt{1-c^2} & c_1\leq c \leq 1
			\end{cases}.
		\end{equation}
	\end{widetext}
	\section{Optimization for the separable case}
	\label{Appendix:separable}
We prove the separable bound by giving an upper bound and then showing that it is saturated. Let
\begin{equation}
	a=\cos\theta,
	\qquad
	b=\sin\theta,
	\qquad
	0\leq b\leq a.
\end{equation}
For product pre- and post-selections, write
\begin{equation}
	\ket{\psi_s}=\ket{\psi_1}\otimes\ket{\psi_2},
	\qquad
	\ket{\phi_s}=\ket{\phi_1}\otimes\ket{\phi_2},
\end{equation}
with
\begin{equation}
	r_j=\abs{\braket{\phi_j}{\psi_j}},
	\qquad
	r_1r_2=c.
\end{equation}
Starting from Eq. (\ref{Eq:optimize-separable}), we introduce the following notation:
\begin{equation}
	X_j=\bra{\phi_j}X\ket{\psi_j},
	\quad
	Y_j=\bra{\phi_j}Y\ket{\psi_j},
	\quad
	Z_j=\bra{\phi_j}Z\ket{\psi_j}.
\end{equation}
Then
\begin{equation}\label{Eq:AppC_transition}
	\bra{\phi_s}B\ket{\psi_s}
	=
	2\left(aX_1X_2+bY_1Y_2\right),
\end{equation}
and therefore
\begin{equation}\label{Eq:AppC_triangle}
	\frac{1}{2}
	\abs{\bra{\phi_s}B\ket{\psi_s}}
	\leq
	a\abs{X_1X_2}+b\abs{Y_1Y_2}.
\end{equation}
For each local pair define the nonnegative quantities
\begin{equation}
	u_j=1-\abs{X_j}^2,
	\qquad
	v_j=1-\abs{Y_j}^2.
\end{equation}
The single-qubit identity
\begin{equation}\label{Eq:AppC_Pauli_identity}
	\abs{X_j}^2+\abs{Y_j}^2+\abs{Z_j}^2
	=
	2-r_j^2
\end{equation}
implies
\begin{equation}\label{Eq:AppC_uv_bound}
	u_j+v_j
	=
	2-\abs{X_j}^2-\abs{Y_j}^2
	\geq
	r_j^2.
\end{equation}
Moreover, by the arithmetic-geometric mean inequality,
\begin{equation}\label{Eq:AppC_X_product_bound}
	\abs{X_1X_2}
	=
	\sqrt{(1-u_1)(1-u_2)}
	\leq
	1-\frac{u_1+u_2}{2},
\end{equation}
and similarly
\begin{equation}\label{Eq:AppC_Y_product_bound}
	\abs{Y_1Y_2}
	\leq
	1-\frac{v_1+v_2}{2}.
\end{equation}
Substituting Eqs. (\ref{Eq:AppC_X_product_bound}) and (\ref{Eq:AppC_Y_product_bound}) into Eq. (\ref{Eq:AppC_triangle}) gives
\begin{align}
	\frac{1}{2}
	\abs{\bra{\phi_s}B\ket{\psi_s}}
	&\leq
	a+b
	-
	\frac{1}{2}
	\left[
	a(u_1+u_2)+b(v_1+v_2)
	\right]
	\nonumber\\
	&\leq
	a+b
	-
	\frac{b}{2}
	\left[
	u_1+v_1+u_2+v_2
	\right]
	\nonumber\\
	&\leq
	a+b
	-
	\frac{b}{2}
	\left(r_1^2+r_2^2\right).
\end{align}
In the second line we used $a\geq b$, and in the third line we used Eq. (\ref{Eq:AppC_uv_bound}). Since $r_1^2+r_2^2\geq2r_1r_2=2c$, we obtain the global upper bound
\begin{equation}\label{Eq:AppC_upper_bound}
	\abs{\bra{\phi_s}B\ket{\psi_s}}
	\leq
	2\left[a+b(1-c)\right].
\end{equation}
Since the preceding inequalities hold for every product pair satisfying $r_1r_2=c$, Eq.~(\ref{Eq:AppC_upper_bound}) is a global upper bound, not merely a bound on the saturating family introduced below.
It remains to show that this upper bound is attainable. Choose $r_1=r_2=\sqrt{c}$. For each side, choose $\ket{\psi_j}$ with Bloch vector
\begin{equation}
	\vb{s}_1=\left(\sqrt{c},0,\sqrt{1-c}\right),
	\qquad
	\vb{s}_2=\left(\sqrt{c},0,-\sqrt{1-c}\right),
\end{equation}
and set
\begin{equation}
	\ket{\phi_j}=X\ket{\psi_j}.
\end{equation}
Then
\begin{equation}
	\abs{\braket{\phi_j}{\psi_j}}
	=
	\abs{\bra{\psi_j}X\ket{\psi_j}}
	=
	\sqrt{c},
\end{equation}
so the total overlap is $c$. Furthermore,
\begin{equation}
	X_j=\bra{\phi_j}X\ket{\psi_j}=1,
	\quad
	Y_j=\bra{\phi_j}Y\ket{\psi_j}=i\bra{\psi_j}Z\ket{\psi_j}.
\end{equation}
Because the two $Z$ components are opposite, this gives
\begin{equation}
	X_1X_2=1,
	\qquad
	Y_1Y_2=1-c.
\end{equation}
Therefore
\begin{equation}
	\abs{\bra{\phi_s}B\ket{\psi_s}}
	=
	2\left[a+b(1-c)\right],
\end{equation}
which saturates Eq. (\ref{Eq:AppC_upper_bound}). Hence
\begin{equation}\label{Eq:AppC_final_bound}
	\twonorm{B}{c,s}
	=
	2\left[\cos\theta+\sin\theta(1-c)\right]
	=
	\lambda_+-\Lambda_-c.
	\end{equation}
\section{Auxiliary optimization for the maximal half-separable bound}
\label{App:MaxHalfSepAux}

We justify the auxiliary optimization step used in the derivation of the maximal half-separable bound.  Let
\begin{equation}
	H_2(k,c)=2\frac{\sqrt{1+2k+2k^2}-kc}{\sqrt{1+k^2}},
	\qquad k\in[0,1],
\end{equation}
and define
\begin{equation}
	f(k)=\frac{1+k-k^2}{\sqrt{1+2k+2k^2}} .
\end{equation}
A direct differentiation gives
\begin{equation}
	\frac{\partial H_2(k,c)}{\partial k}
	=\frac{2\left[f(k)-c\right]}{(1+k^2)^{3/2}} .
\end{equation}
Thus the stationary condition for the second branch is equivalent to
\begin{equation}
	f(k)=c .
\end{equation}
Moreover,
\begin{equation}
	f'(k)=
	-\frac{k(2k^2+3k+3)}{(1+2k+2k^2)^{3/2}}\leq 0,
\end{equation}
with equality only at $k=0$. Hence $f$ is strictly decreasing on $(0,1]$. Since
\begin{equation}
	f(0)=1,
	\qquad
	f(1)=\frac{1}{\sqrt{5}},
\end{equation}
there is, for every $c\in[1/\sqrt{5},1]$, a unique solution
\begin{equation}
	\kappa(c)\in[0,1]
\end{equation}
to the stationary equation. In particular,
\begin{equation}
	\kappa(1/\sqrt{5})=1,
	\qquad
	\kappa(1)=0 .
\end{equation}
Since $f(k)>c$ for $k<\kappa(c)$ and $f(k)<c$ for $k>\kappa(c)$, this stationary point is the global maximum of the second branch on $k\in[0,1]$.

It remains to check branch consistency.  At the solution $c=f(k)$, and using
\begin{equation}
	c_0(k)=\frac{k}{\sqrt{1+2k+2k^2}},
	\qquad
	c_1(k)=\frac{1+k}{\sqrt{1+2k+2k^2}},
\end{equation}
we have
\begin{equation}
	c-c_0(k)=\frac{1-k^2}{\sqrt{1+2k+2k^2}}\geq0,
\end{equation}
and
\begin{equation}
	c_1(k)-c=\frac{k^2}{\sqrt{1+2k+2k^2}}\geq0 .
\end{equation}
Therefore
\begin{equation}
	c_0(k)\leq c\leq c_1(k),
\end{equation}
so the maximizing point indeed lies on the second branch. Finally, since $H_2(0,c)=2$, the optimized second branch obeys
\begin{equation}
	H_2(\kappa(c),c)\geq 2 .
\end{equation}
Together with the Cauchy--Schwarz bound on the third branch used in the main text, this completes the comparison between the second and third branches.	
	\section{Standard detection}
	\label{Appendix:standard_detection}
	Here we analyze the outcome of a standard Bell test for the entangled states that saturate the Tsirelson and half-separable bounds derived in Sec. \ref{Sec:CHSH_bounds_in_PPS_settings-General_Bounds}.
	\\
	We recall that for any Bell operator of the form given by 
	Eqs.~(\ref{Eq:Local_Equivalent_B}) and~(\ref{gen_bell1})
	the threshold for entanglement detection is 
	\begin{equation}
		\abs{\bra{\psi}B\ket{\psi}}>\beta_{\text{sep}}(B)=\Lambda_+=2\cos(\theta).
	\end{equation}
Thus, in a standard CHSH entanglement test, entanglement is certified by surpassing the standard separable bound for the chosen Bell operator.
	\subsection{Tsirelson}
	To investigate standard entanglement detection with respect to the states used for the two-state Tsirelson bound, we may use either of the states given by Eq. (\ref{Eq:Tsirelson_states}) and get
	\begin{equation}\label{Eq:S_1}
		\mathcal{S}_1:=
		\abs{\bra{\psi}B\ket{\psi}}
		=
		\abs{\bra{\phi}B\ket{\phi}}
		=
		2c\left(
		\cos(\theta)+\sin(\theta)
		\right),	
	\end{equation}
	which is a linear function with respect to the overlap.
	These states are certified by the standard CHSH test only above a threshold value of $c$, namely
	\begin{equation}
		c_{\text{threshold}}^{(1)}
		<c\leq 1,
	\end{equation}
	where the overlap threshold for detection is given by
	\begin{equation}
		c_{\text{threshold}}^{(1)}
		=\left(1+\frac{\Lambda_-}{\Lambda_+}\right)^{-1}=\left(1+\tan(\theta)\right)^{-1}.
	\end{equation}
	\subsection{Half-separable}
	To investigate standard entanglement detection with respect to a representative $\ket{\psi}$ --- the entangled state that optimizes the half-separable bound $\twonorm{B}{c,1s}$, we recall that
	\begin{equation}
		\ket{\psi}=c\ket{\phi_s}
		+
		\sqrt{1-c^2}\ket{\phi_{s,\perp}^{(B)}}.
	\end{equation}
	The orthogonal state is given by
	\begin{equation}
		\ket{\phi_{s,\perp}^{(B)}}=
		\frac{
			B-\expval{B}_{\phi_s}
		}{\Delta\left(B\right)_{\phi_s}}\ket{\phi_s}.
	\end{equation}
	Therefore, we get
	\begin{equation}
		\begin{aligned}
			&\bra{\psi}B\ket{\psi}
			\\&=
			\left(
			c\bra{\phi_s}
			+
			\sqrt{1-c^2}\bra{\phi_{s,\perp}^{(B)}}
			\right)B
			\left(c\ket{\phi_s}
			+
			\sqrt{1-c^2}\ket{\phi_{s,\perp}^{(B)}}\right)
			\\&=
			c^2\expval{B}_{\phi_s}
			+c\sqrt{1-c^2}
			\left(
			\bra{\phi_s}
			B
			\ket{\phi_{s,\perp}^{(B)}}
			+
			\bra{\phi_{s,\perp}^{(B)}}
			B
			\ket{\phi_s}
			\right)
			\\&
			+(1-c^2)\expval{B}_{\phi_{s,\perp}^{(B)}}
			\\&=
			c^2\expval{B}_{\phi_s}+2c\sqrt{1-c^2}\Delta\left(B\right)_{\phi_s}+(1-c^2)\expval{B}_{\phi_{s,\perp}^{(B)}}.
		\end{aligned}	
	\end{equation}
	Next, we expand $\expval{B}_{\phi_{s,\perp}^{(B)}}$ using $\ket{\phi_s}$ and get
	\begin{equation}
		\begin{aligned}
			\expval{B}_{\phi_{s,\perp}^{(B)}}&=
			\frac{\expval{(B-\expval{B}_{\phi_s})B(B-\expval{B}_{\phi_s})}_{\phi_s}}{\Delta^2\left(B\right)_{\phi_s}}
			\\&=
			\frac{
				\expval{B^3}_{\phi_s}-2\expval{B^2}_{\phi_s}\expval{B}_{\phi_s}+\expval{B}_{\phi_s}^3
			}
			{\Delta^2\left(B\right)_{\phi_s}}.
		\end{aligned}	
	\end{equation}
	To proceed, we may use Eq. (\ref{gen_bell2}) along with
	$\expval{X\otimes X}_{\phi_s}=x$, $\expval{Y\otimes Y}_{\phi_s}=0$, $\expval{Z\otimes Z}_{\phi_s}=-(1-x)$ and get
	\begin{equation}
		\begin{aligned}
			&\bra{\psi}B\ket{\psi}
			=
			c^2\expval{B}_{\phi_s}
			+2c\sqrt{1-c^2}
			\Delta\left(B\right)_{\phi_s}
			\\&
			+\frac{1-c^2}{\Delta^2\left(B\right)_{\phi_s}}
			\left(
			\expval{B^3}_{\phi_s}-2\expval{B^2}_{\phi_s}\expval{B}_{\phi_s}+\expval{B}_{\phi_s}^3
			\right)	.
		\end{aligned}
	\end{equation}
	with
	\begin{equation}
		\begin{aligned}
			&
			\begin{aligned}
				&\expval{B}_{\phi_s}=2\cos(\theta)x,
				&\expval{B^2}_{\phi_s}=4\left(1+\sin(2\theta)(1-x)\right),
			\end{aligned}
			\\&
			\begin{aligned}
				&\expval{B^3}_{\phi_s}
				=4\left(3\cos(\theta)-\cos(3\theta)\right)x,
			\end{aligned}
		\end{aligned}	
	\end{equation}
	where $x$ is given by Eq. (\ref{Eq:x}).
	Interestingly, this expression can also be written in the following way
	\begin{equation}\label{Eq:S_2}
		\begin{aligned}
			&\mathcal{S}_2:=\abs{\bra{\psi}B\ket{\psi}}=
			\\&\small
			\begin{cases}
				4c\sqrt{1-c^2}(\cos(\theta)+\sin(\theta)) & 0\leq c \leq c_0\\
				4c_0 c_1(\sqrt{1+\sin(2\theta)})+2\cos(\theta)
				\frac{(1-2c_0 c_1)}{c_1-c_0}(c-c_0) 
				& c_0\leq c \leq c_1 \\
				2(\cos(\theta)+2c\sqrt{1-c^2}\sin(\theta)) & c_1\leq c \leq 1
			\end{cases}	
		\end{aligned}
	\end{equation}
	where the coefficients $c_0$ and $c_1$ are given by Eq. (\ref{Eq:c-coefficients}).
	By analyzing these branches, it can be shown that
	the overlap threshold for detection is given by
	\begin{equation}
		c_{\text{threshold}}^{(2)}(\theta)=
		\begin{cases}
			\frac{\left(1+\tan(\theta)\right)(1-2\tan[2](\theta))}{\sqrt{2\tan[2](\theta)+2\tan(\theta)+1}}
			& 0< \theta \leq \theta^*\\
			\sqrt{\frac{1}{2}\left(
				1-
				\sqrt{
					\frac{\sin(2\theta)+\sin[2](\theta)}{1+\sin(2\theta)}	
				}
				\right)} & \theta^*\leq \theta \leq \frac{\pi}{4}
		\end{cases},	
	\end{equation}
	where $\tan(\theta^*)$ is given by the solution of 
	\begin{equation}
		2\tan[3](\theta)+2\tan[2](\theta)-1=0.
	\end{equation}
	The solution gives 
	\begin{equation}
		\begin{aligned}
			\tan(\theta^*)&=\frac{1}{6}\left[
			\left(
			46+6\sqrt{57}
			\right)^{1/3}+
			\left(
			46-6\sqrt{57}
			\right)^{1/3}
			-2
			\right]
			\\&\approx 0.565198.
		\end{aligned}	
	\end{equation}
 Thus the representative entangled state $\ket{\psi}$ is certified by the standard test only for
	\begin{equation}
		c_{\text{threshold}}^{(2)}
		<c<1.
	\end{equation}
	\section{Global optimality of the state-adapted $\gamma=1$ construction}
	\label{Appendix:Global_optimality}
	We show that, for a pure two-qubit source state, the Schmidt-aligned construction maximizes the transition-amplitude witness margin over all local CHSH frames and all product post-selections. By local unitary rotations, the source state may be written in Schmidt form as
	\begin{equation}
		\ket{\psi}
		=
		\alpha\ket{00}
		+
		\beta\ket{11},
		\qquad
		\alpha\geq\beta>0,
		\qquad
		\alpha^2+\beta^2=1.
		\label{Eq:global_opt_schmidt_state}
	\end{equation}
	We denote by
	\begin{equation}
		\sigma_{\vb{n}}:=\vb{n}\cdot\vb{\sigma}
	\end{equation}
	the Pauli observable along the Bloch-sphere direction $\vb{n}$. For a fixed value of the CHSH incompatibility parameter, an arbitrary locally rotated canonical CHSH operator can be written as
	\begin{equation}
		B_{\theta,\mathcal F}
		=
		2K_{\theta,\mathcal F},
		\label{Eq:global_opt_B_equals_2K}
	\end{equation}
	where
	\begin{equation}
		K_{\theta,\mathcal F}
		=
		a\,\sigma_{\vb{u}_{A}}\otimes\sigma_{\vb{u}_{B}}
		+
		b\,\sigma_{\vb{v}_{A}}\otimes\sigma_{\vb{v}_{B}},
		\qquad
		a=\cos\theta,
		\qquad
		b=\sin\theta.
		\label{Eq:global_opt_Ktheta}
	\end{equation}
	Here $0\leq\theta\leq\pi/4$, so that $0\leq b\leq a$, and
	\begin{equation}
		\gamma=\sin(2\theta).
	\end{equation}
	The frame $\mathcal F$ denotes the choice of local orthonormal Bloch-sphere directions,
	\begin{equation}
		\vb{u}_{A}\cdot\vb{v}_{A}=0,
		\qquad
		\vb{u}_{B}\cdot\vb{v}_{B}=0.
		\label{Eq:global_opt_orthogonal_directions}
	\end{equation}
	Thus Eq. (\ref{Eq:global_opt_Ktheta}) represents an arbitrary CHSH operator with fixed $\theta$, equivalently fixed $\gamma$, up to local unitary rotations.
	
	Let $\ket{\phi_s}$ be an arbitrary normalized product post-selection and define its overlap with the source state by
	\begin{equation}
		c=\abs{\braket{\phi_s}{\psi}}.
		\label{Eq:global_opt_overlap}
	\end{equation}
	The transition-amplitude witness margin for the choice $(B_{\theta,\mathcal F},\ket{\phi_s})$ is
	\begin{equation}
		\mathcal M_{\theta,\mathcal F}(\phi_s)
		=
		\abs{\bra{\phi_s}B_{\theta,\mathcal F}\ket{\psi}}
		-
		\twonorm{B_{\theta,\mathcal F}}{c,s}.
		\label{Eq:global_opt_margin_definition}
	\end{equation}
	Using the fixed-operator separable PPS bound,
	\begin{equation}
		\twonorm{B_{\theta,\mathcal F}}{c,s}
		=
		2\left[a+b(1-c)\right],
		\label{Eq:global_opt_separable_bound}
	\end{equation}
	we obtain
	\begin{align}
		\mathcal M_{\theta,\mathcal F}(\phi_s)
		&=
		2\abs{\bra{\phi_s}K_{\theta,\mathcal F}\ket{\psi}}
		-
		2\left[a+b(1-c)\right]
		\nonumber\\
		&=
		2\left[
		\abs{\bra{\phi_s}K_{\theta,\mathcal F}\ket{\psi}}
		+
		bc
		-
		a
		-
		b
		\right].
		\label{Eq:global_opt_margin_rewritten}
	\end{align}
	Therefore, to upper-bound the margin, it is enough to upper-bound
	\begin{equation}
		\abs{\bra{\phi_s}K_{\theta,\mathcal F}\ket{\psi}}
		+
		b\abs{\braket{\phi_s}{\psi}}.
	\end{equation}
	
	We claim that, for every product post-selection $\ket{\phi_s}$ and every local CHSH frame $\mathcal F$,
	\begin{equation}
		\abs{\bra{\phi_s}K_{\theta,\mathcal F}\ket{\psi}}
		+
		b\abs{\braket{\phi_s}{\psi}}
		\leq
		a\alpha+b(\alpha+\beta).
		\label{Eq:global_opt_key_bound}
	\end{equation}
	To prove this, define
	\begin{equation}
		x=\bra{\phi_s}K_{\theta,\mathcal F}\ket{\psi},
		\qquad
		y=\braket{\phi_s}{\psi}.
		\label{Eq:global_opt_xy_definitions}
	\end{equation}
	Choose phases $\chi$ and $\xi$ such that
	\begin{equation}
		e^{-i\chi}x=\abs{x},
		\qquad
		e^{-i\xi}y=\abs{y}.
		\label{Eq:global_opt_phase_choice}
	\end{equation}
	If either $x$ or $y$ vanishes, the corresponding phase may be chosen arbitrarily. Now define the generally unnormalized two-qubit vector
	\begin{equation}
		\ket{\eta}
		=
		e^{i\chi}K_{\theta,\mathcal F}\ket{\phi_s}
		+
		b\,e^{i\xi}\ket{\phi_s}.
		\label{Eq:global_opt_eta_definition}
	\end{equation}
	Since $K_{\theta,\mathcal F}$ is Hermitian, we have
	\begin{align}
		\braket{\eta}{\psi}
		&=
		e^{-i\chi}\bra{\phi_s}K_{\theta,\mathcal F}\ket{\psi}
		+
		b\,e^{-i\xi}\braket{\phi_s}{\psi}
		\nonumber\\
		&=
		\abs{x}+b\abs{y}.
	\end{align}
	Thus
	\begin{equation}
		\abs{\bra{\phi_s}K_{\theta,\mathcal F}\ket{\psi}}
		+
		b\abs{\braket{\phi_s}{\psi}}
		=
		\abs{\braket{\eta}{\psi}}.
		\label{Eq:global_opt_eta_identity}
	\end{equation}
	
	Let $\mu_1\geq\mu_2\geq0$ be the Schmidt coefficients of $\ket{\eta}$. Since the Schmidt coefficients of $\ket{\psi}$ are $\alpha$ and $\beta$, the overlap between $\ket{\eta}$ and $\ket{\psi}$ is bounded by
	\begin{equation}
		\abs{\braket{\eta}{\psi}}
		\leq
		\alpha\mu_1+\beta\mu_2.
		\label{Eq:global_opt_schmidt_overlap_bound}
	\end{equation}
	This follows from the singular-value overlap bound. Equivalently, writing bipartite vectors as $2\times2$ coefficient matrices, the Schmidt coefficients are the singular values of the corresponding coefficient matrix, and Eq. (\ref{Eq:global_opt_schmidt_overlap_bound}) is an application of von Neumann's trace inequality.
	
	We now bound $\mu_1$ and $\mu_1+\mu_2$. First, the largest Schmidt coefficient of $\ket{\eta}$ is its maximal overlap with a normalized product state:
	\begin{equation}
		\mu_1
		=
		\sup_{\ket{\chi_s}\in\Prod}
		\abs{\braket{\chi_s}{\eta}}.
		\label{Eq:global_opt_largest_schmidt_product_overlap}
	\end{equation}
	For an arbitrary normalized product state $\ket{\chi_s}$, define
	\begin{equation}
		d=\abs{\braket{\chi_s}{\phi_s}}.
		\label{Eq:global_opt_d_definition}
	\end{equation}
	Using Eq. (\ref{Eq:global_opt_eta_definition}) and the triangle inequality,
	\begin{align}
		\abs{\braket{\chi_s}{\eta}}
		&\leq
		\abs{\bra{\chi_s}K_{\theta,\mathcal F}\ket{\phi_s}}
		+
		b\abs{\braket{\chi_s}{\phi_s}}
		\nonumber\\
		&=
		\abs{\bra{\chi_s}K_{\theta,\mathcal F}\ket{\phi_s}}
		+
		bd.
		\label{Eq:global_opt_product_overlap_eta_bound}
	\end{align}
	Since both $\ket{\chi_s}$ and $\ket{\phi_s}$ are product states with overlap $d$, the separable PPS bound for $K_{\theta,\mathcal F}=B_{\theta,\mathcal F}/2$ gives
	\begin{equation}
		\abs{\bra{\chi_s}K_{\theta,\mathcal F}\ket{\phi_s}}
		\leq
		a+b(1-d).
		\label{Eq:global_opt_K_separable_bound}
	\end{equation}
	Substituting Eq. (\ref{Eq:global_opt_K_separable_bound}) into Eq. (\ref{Eq:global_opt_product_overlap_eta_bound}) gives
	\begin{equation}
		\abs{\braket{\chi_s}{\eta}}
		\leq
		a+b(1-d)+bd
		=
		a+b.
	\end{equation}
	Taking the supremum over all normalized product states $\ket{\chi_s}$ therefore yields
	\begin{equation}
		\mu_1\leq a+b.
		\label{Eq:global_opt_mu1_bound}
	\end{equation}
	
	Second, we bound the sum of Schmidt coefficients. The three vectors
	\begin{equation}
		\sigma_{\vb{u}_{A}}\otimes\sigma_{\vb{u}_{B}}\ket{\phi_s},
		\qquad
		\sigma_{\vb{v}_{A}}\otimes\sigma_{\vb{v}_{B}}\ket{\phi_s},
		\qquad
		\ket{\phi_s}
	\end{equation}
	are all normalized product vectors, because local Pauli operators map product states to product states. Therefore Eq. (\ref{Eq:global_opt_eta_definition}) expresses $\ket{\eta}$ as a linear combination of three normalized product vectors with coefficient magnitudes $a$, $b$, and $b$. The sum of Schmidt coefficients is subadditive under vector addition, and the sum of Schmidt coefficients of a normalized product vector is one. 
	Equivalently, after representing bipartite vectors by coefficient matrices, the sum of Schmidt coefficients is the trace norm, and the trace norm obeys the triangle inequality.
	Hence
	\begin{equation}
		\mu_1+\mu_2
		\leq
		a+b+b
		=
		a+2b.
		\label{Eq:global_opt_mu_sum_bound}
	\end{equation}
	
	Combining Eqs. (\ref{Eq:global_opt_schmidt_overlap_bound}), (\ref{Eq:global_opt_mu1_bound}), and (\ref{Eq:global_opt_mu_sum_bound}), and using $\alpha\geq\beta$, we obtain
	\begin{align}
		\abs{\braket{\eta}{\psi}}
		&\leq
		\alpha\mu_1+\beta\mu_2
		\nonumber\\
		&=
		(\alpha-\beta)\mu_1+\beta(\mu_1+\mu_2)
		\nonumber\\
		&\leq
		(\alpha-\beta)(a+b)+\beta(a+2b)
		\nonumber\\
		&=
		a\alpha+b(\alpha+\beta).
		\label{Eq:global_opt_eta_final_bound}
	\end{align}
	Together with Eq. (\ref{Eq:global_opt_eta_identity}), this proves Eq. (\ref{Eq:global_opt_key_bound}).
	
	Substituting Eq. (\ref{Eq:global_opt_key_bound}) into Eq. (\ref{Eq:global_opt_margin_rewritten}) gives
	\begin{align}
		\mathcal M_{\theta,\mathcal F}(\phi_s)
		&=
		2\left[
		\abs{\bra{\phi_s}K_{\theta,\mathcal F}\ket{\psi}}
		+
		bc
		-
		a
		-
		b
		\right]
		\nonumber\\
		&\leq
		2\left[
		a\alpha+b(\alpha+\beta)-a-b
		\right]
		\nonumber\\
		&=
		2\left[
		(\alpha-1)\cos\theta
		+
		(\alpha+\beta-1)\sin\theta
		\right].
		\label{Eq:global_opt_margin_upper_bound}
	\end{align}
	Thus, for fixed $\theta$, no local CHSH frame and no product post-selection can exceed the right-hand side of Eq. (\ref{Eq:global_opt_margin_upper_bound}).
	
	We now show that this upper bound is saturated. Choose the Schmidt-aligned product post-selection
	\begin{equation}
		\ket{\phi_s}=\ket{00}
	\end{equation}
	and the Schmidt-aligned CHSH operator
	\begin{equation}
		B_{\theta}^{\mathrm{Sch}}
		=
		2\left(
		\cos\theta\,Z\otimes Z
		+
		\sin\theta\,X\otimes X
		\right).
		\label{Eq:global_opt_schmidt_aligned_B}
	\end{equation}
	For this choice,
	\begin{equation}
		c=\abs{\braket{00}{\psi}}=\alpha,
	\end{equation}
	and
	\begin{equation}
		\abs{\bra{00}B_{\theta}^{\mathrm{Sch}}\ket{\psi}}
		=
		2(\alpha\cos\theta+\beta\sin\theta).
		\label{Eq:global_opt_schmidt_transition}
	\end{equation}
	Using the separable PPS bound in Eq. (\ref{Eq:global_opt_separable_bound}), the margin becomes
	\begin{align}
		\mathcal M_{\theta}^{\mathrm{Sch}}
		&=
		2(\alpha\cos\theta+\beta\sin\theta)
		-
		2\left[
		\cos\theta+\sin\theta(1-\alpha)
		\right]
		\nonumber\\
		&=
		2\left[
		(\alpha-1)\cos\theta
		+
		(\alpha+\beta-1)\sin\theta
		\right].
		\label{Eq:global_opt_schmidt_margin}
	\end{align}
	This equals the upper bound in Eq. (\ref{Eq:global_opt_margin_upper_bound}). Hence, for every fixed $\theta$,
	\begin{equation}
		\mathcal M_{\max}(\theta)
		=
		2\left[
		(\alpha-1)\cos\theta
		+
		(\alpha+\beta-1)\sin\theta
		\right],
		\label{Eq:global_opt_Mmax_theta}
	\end{equation}
	where the maximum is taken over all local CHSH frames and all product post-selections, with each post-selection compared at its corresponding overlap $c=\abs{\braket{\phi_s}{\psi}}$.
	
	It remains to optimize Eq. (\ref{Eq:global_opt_Mmax_theta}) over $\theta$. Differentiating gives
	\begin{equation}
		\frac{d\mathcal M_{\max}}{d\theta}
		=
		2\left[
		(1-\alpha)\sin\theta
		+
		(\alpha+\beta-1)\cos\theta
		\right].
		\label{Eq:global_opt_derivative}
	\end{equation}
	For an entangled state, $\beta>0$, and hence $\alpha<1$. Moreover,
	\begin{equation}
		(\alpha+\beta)^2
		=
		\alpha^2+\beta^2+2\alpha\beta
		=
		1+2\alpha\beta
		>
		1,
	\end{equation}
	so
	\begin{equation}
		\alpha+\beta>1.
	\end{equation}
	Since $\sin\theta,\cos\theta\geq0$ for $\theta\in[0,\pi/4]$, Eq. (\ref{Eq:global_opt_derivative}) implies
	\begin{equation}
		\frac{d\mathcal M_{\max}}{d\theta}>0
	\end{equation}
	throughout the interval. 
	Therefore the maximum is attained at $\theta=\frac{\pi}{4}$, equivalently at $\gamma=1$.
	
	Consequently, for every pure two-qubit source state with nonzero concurrence, the state-adapted transition-amplitude witness margin optimized over all local CHSH frames and all product post-selections is maximized by a maximally incompatible CHSH operator with $\gamma=1$ and by the Schmidt-aligned product post-selection.
	
	We emphasize that this statement concerns the state-adapted optimization in which each product post-selection is evaluated at its own overlap $c=\abs{\braket{\phi_s}{\psi}}$. It does not by itself prove optimality at an externally prescribed overlap $c_0$ that is held fixed independently of the post-selection.
		\bibliographystyle{apsrev4-2}	
	\bibliography{WV_CHSH}	
	
\end{document}